\documentclass[a4paper,12pt]{article}

\usepackage[utf8x]{inputenc}
\usepackage{graphicx}
\usepackage{overcite}
\usepackage{amsmath}
\usepackage{amssymb}
\usepackage{float}
\usepackage{rotating}
\usepackage{longtable}
\usepackage{scrextend}
\usepackage[font=small,labelfont=bf]{caption}
\usepackage{url}

\newcommand{\onlinecite}[1]{\hspace{-1 ex} \nocite{#1}\citenum{#1}}

\title{Context-Driven Exploration of Complex Chemical Reaction Networks}
\author{Gregor N.\ Simm and Markus Reiher\thanks{corresponding author: markus.reiher@phys.chem.ethz.ch; Phone: +41446334308; Fax: +41446331594}
\vspace{10 mm}\\
ETH Z\"urich, Laboratory of Physical Chemistry, \\ Vladimir-Prelog-Weg 2, 8093 Z\"urich, Switzerland.
}

\begin{document}

\maketitle

\section*{Abstract}

The construction of a reaction network containing all relevant intermediates and elementary reactions is necessary for the accurate description of chemical processes.
In the case of a complex chemical reaction (involving, for instance, many reactants or highly reactive species), the size of such network may grow rapidly.
Here, we present a computational protocol that constructs such reaction networks in a fully automated fashion steered in an
intuitive, graph-based fashion through a single graphical user interface.
Starting from a set of initial reagents new intermediates are explored through intra- and intermolecular reactions of
already explored intermediates or new reactants presented to the network.
This is done by assembling reactive complexes based on heuristic rules derived from conceptual electronic-structure theory
and exploring the corresponding approximate reaction path.
A subsequent path refinement leads to a minimum-energy path which connects the new intermediate to the existing ones to form a connected reaction network.
Tree traversal algorithms are then employed to detect reaction channels and catalytic cycles.
We apply our protocol to the formose reaction to study different pathways of sugar formation
and to rationalize its autocatalytic nature.

\section{Introduction}

Complex reaction mechanisms, in which multiple reaction pathways compete, can be found in many areas of chemistry including transition-metal catalysis,\cite{Masters2011}
polymerization,\cite{Vinu2012} and atmospheric chemistry.\cite{Vereecken2015}
While many reactions in chemical synthesis result in the selective formation of a dominant product, the presence of many reactants or highly reactive species can lead to the generation of multiple products.
A detailed understanding of all reaction pathways would allow one to study the evolution of a system over time given a set of initial conditions
(e.g., reactants and their concentration, temperature, and pressure)
and provide means to effectively perturb the chemical system, and hence, to control product composition.
A complete understanding of a reaction network and its alchemically derived relatives, which may consider derivatives of the original reactants, is crucial for molecular compound design and for avoiding undesired side reactions (such as molecule decomposition).

The description of a chemical process in such detail requires the knowledge of all possible chemical compounds that can be formed in this process under the given conditions
and the elementary reactions connecting them.
For all but the simplest systems, exhaustive manual explorations are unfeasible, and therefore, manual studies are limited to the expected dominant reaction paths.
In addition, such manual explorations are slow, tedious, and error-prone.

There exist already many approaches for automated explorations of configuration spaces producing both intermediates and transition states.
For example, in reactive molecular-dynamics simulations, the nuclear equations of motion are solved to explore and sample the part of configuration space
that is accessible under the constraints imposed by a pre-defined thermodynamic ensemble.
Recently, such approaches were successfully applied to study prebiotic reactions occurring in the Urey--Miller experiment.\cite{Saitta2014,Wang2014,Wang2016}

As the configuration space can become very large, comprising multiple copies of all chemical species involved in the reaction,
computational costs of carrying out first-principles calculations grow rapidly.
This issue can be overcome by the application of a reactive force-field.\cite{vanDuin2001,Dontgen2015}
Unfortunately, next to the reduced accuracy, force-field parameters will, in general, not be available for any type of system which limits their applicability.
Therefore, hybrid quantum-mechanical--molecular-mechanical approaches have been frequently applied to explore different
reaction paths of complex systems with many degrees of freedom such as enzymatic reactions
(for examples see Refs.~\onlinecite{Fischer1992,Florian2003,Garcia-Viloca2004,Imhof2009,Reidelbach2016,Imhof2016} and reviews by Senn and Thiel~\cite{Senn2006,Senn2007,Senn2009}).
However, to increase the possibility of a reaction to occur, the temperature and pressure of the simulation need to be increased greatly,
which in turn reduces the accuracy of the sampling statistics.

By contrast, global reaction route mapping (GRRM)\cite{Ohno2004,Ohno2008,Maeda2013,Satoh2015,Satoh2016,Ohno2017} employs a stationary approach, where the Born--Oppenheimer potential-energy surface (PES), starting from one initial configuration,
is explored based on local curvature information.
While being highly systematic, this approach is difficult to apply for the exploration of truly large reactive systems.

A complementary strategy is to apply conceptual knowledge of chemistry to explore reaction mechanisms efficiently.
In particular in organic chemistry, reactivity is often well captured by a set of heuristic rules, commonly denoted as ``arrow pushing''.\cite{Levy2017}
Graph-based rules originating from concepts of bond order and valence are prevalent.\cite{Rucker2004,Todd2005,Chen2008,Chen2009,Graulich2010,Kayala2011,Kayala2012,Zimmerman2013a,Rappoport2014,Zimmerman2015,Zubarev2015,Habershon2015,Habershon2016}
By applying these transformation rules to graph representations of molecules new chemical species, that are energetically accessible, are generated.
For example, Zimmerman\cite{Zimmerman2013a,Zimmerman2015,Pendleton2016,Ludwig2016,Dewyer2017} applied this approach to several reactions in organic and organometallic chemistry.
Aspuru-Guzik and co-workers\cite{Rappoport2014,Zubarev2015} developed a methodology based on formal bond orders and Hammond's postulate to model
prebiotic reactions such as the \textit{formose} reaction\cite{Butlerow1861} and to estimate their activation barriers.

In the afore-mentioned studies, quantum chemical methods were applied to perform structure optimizations and transition-state searches,
whereas the exploration of new intermediates was accomplished by employing a graph representation of the molecules.
While being computationally efficient, there are some potential drawbacks to this approach.
Graph-based rules that rely on the concept of valence perform well for many organic molecules, but may fail for systems containing species
with complex electronic structures such as transition-metal clusters.
In addition, to ensure an exhaustive exploration with such an approach, completeness of the set of transformation rules is required.
However, for an arbitrary, unknown chemical system this cannot be guaranteed.
One will then be restricted to known chemical transformations, which may hamper the discovery of new chemical processes.

In 2015, we introduced a less concept-driven method\cite{Bergeler2015} based on system-independent heuristic rules derived from electronic structure.
Employing a scalar-field descriptor such as the electron localization function\cite{Becke1990} we identified \textit{reactive sites} in a chemical system.
Two reactive sites are then brought into close proximity resembling a reactive complex of high energy.
After standard structure-optimization, new intermediates were discovered.
Subsequently, based on a root-mean-square deviation criterion, pairs of structures that may be interconverted by one elementary reaction were identified.
For such a pair of structures, standard methods were employed to find a transition state and to verify it through an intrinsic reaction coordinate (IRC) calculation.\cite{Fukui1970}

In that study,\cite{Bergeler2015} the reactivity of Schrock's nitrogen-fixating molybdenum catalyst\cite{Yandulov2003, Yandulov2003a} under acidic and reducing conditions was investigated.
Several approximations were made in the exploration, which are overcome in the present study.
Firstly, reactive complexes were generated only between the catalyst and a proton (originating from an acid) -- reactions between new intermediates,
decomposition reactions, and reactions with small molecules (e.g., solvent and ligand molecules) were neglected.
Secondly, since one reactive species was just a single atom, the relative orientation of species forming a reactive complex needed not to be considered.
Finally, the reactants' conformational degrees of freedom were not explored.

In this work, we present a new, generalized approach called \texttt{Chemoton}
(named after a theory for the functioning of living systems proposed by G\'anti\cite{Ganti1975}) for the exploration of arbitrary molecular systems, which
requires the introduction of an appropriate descriptor for the identification of reactive sites.
To ensure the uniqueness of intermediates throughout the exploration, a graph theoretical approach is employed.
In addition, conformers are generated for each unique intermediate.
To allow for an exhaustive exploration, not only reactions between intermediates but also intramolecular reactions are studied.
Furthermore, reactive complexes are assembled under consideration of the relative orientation of the reactive species.
All stationary points found on the different PESs are refined by structure optimization and IRC calculations. All of this is carried out
fully automatically through a single graphical user interface steered by a computer mouse.
Finally, to summarize the kinetic and thermodynamic data gathered in the exploration,
results are visualized by an automatically generated network graph.

To illustrate the functionality of our machinery, we apply it to the \textit{formose} reaction,
a well-studied prebiotic oligomerization reaction of formaldehyde resulting in a highly complex mixture of linear and branched compounds, including monosaccharides.\cite{Butlerow1861,Eschenmoser1992,Delidovich2014}
The identification of all products poses a major experimental challenge and the exact
composition has not been elucidated yet, although over 50 products have already been characterized.\cite{Decker1982,Zweckmair2014}
While some major reaction pathways are known,\cite{Breslow1959,Bissette2013} many mechanistic details are not.\cite{Kim2011}
Due to the formation of biologically important monosaccharides, the formose reaction may constitute a plausible
prebiotic source of sugars.
In recent work\cite{Proppe2016}, we studied the effect of uncertainties in calculated free energies on quantitative fluxes at the example of a tiny sub-network of the formose reaction,
which already featured many conceptual challenges of the full network.
The network generated in the present study is now vastly extended and can be considered the first step toward a complete understanding of this reaction.

\section{Ingredients of Reaction Explorations by \texttt{Chemoton}}
\label{sec:protocol}

Starting from a set of initial conditions, our exploration protocol to be detailed in this section is applied repeatedly to expand a reaction network in a rolling fashion.
These conditions comprise the reactants and their concentrations, solvents, and standard thermodynamic ensemble parameters such as temperature and pressure.
In addition, the time scale of the reaction is relevant as it allows one to define a slowest
reaction which still affects the concentration of all species in a significant manner (for details see Refs.~\onlinecite{Bergeler2015,Proppe2016}).
Reactions slower than that one can be safely discarded, whereas reactions which are much faster can be considered to be in quasi-equilibrium.\cite{Proppe2016}

Unlike reactive molecular-dynamics simulations, a concept-driven exploration does not take place on a single PES
but on multiple low-dimensional PESs consisting of rather few nuclear coordinates that can represent specific elementary
reactions as will be discussed in Section \ref{subsec:reactive_complex} below.
Exploring many low-dimensional PESs instead of one of very high dimension bears several advantages.
Calculations are, in general, faster due to the reduced number of atoms.
Geometry optimizations and transition-state searches converge more quickly
and the exploration can be more easily steered into regions of interest.

\subsection{Generation of Conformers}
\label{subsec:conformers}

A PES is explored starting from one minimum-energy structure which may be a molecule or a cluster of molecules
(such as a microsolvated solute).
Usually, there exist many minima which can be reached from this minimum through a series of elementary reactions
featuring a sufficiently low reaction barrier.
Often, these minima will turn out to be conformers of the same molecular configuration.
With the introduction of some electronic-structure measure for molecular bonds,
these minima can be explored very efficiently employing conformer generators.\cite{Vainio2007,Leite2007,O'Boyle2011,Miteva2010,Riniker2015}
We determine the molecular bonding by calculating Mayer bond orders from an electronic wave function.\cite{Mayer1983}
The geometries of the generated conformers are subsequently optimized with quantum chemical methods to obtain sufficiently reliable minimum structures.

If the time scale of a reaction can be assumed to be longer than the time the conformers require for equilibration,
transition states between the conformers do not need to be optimized, which reduces the the computational effort significantly.
In this case, at a given temperature, only a fraction of the conformers can be assumed to be significantly populated, and hence,
only this fraction needs to be considered in the subsequent steps of the exploration protocol.
Otherwise, transition states need to be located (see Section~\ref{subsec:mep}).

\subsection{Assembly of Reactive Complexes and Induction of Reactions}
\label{subsec:reactive_complex}

Searching for minima that can only be reached by overcoming a non-negligible barrier is not straightforward,
as, in general, this requires a chemical transformation (i.e., breaking or forming bonds).
In the following steps of our protocol, we distinguish between \textit{intermolecular} and \textit{intramolecular} reactions.

\begin{figure}[H]
\centering
\includegraphics{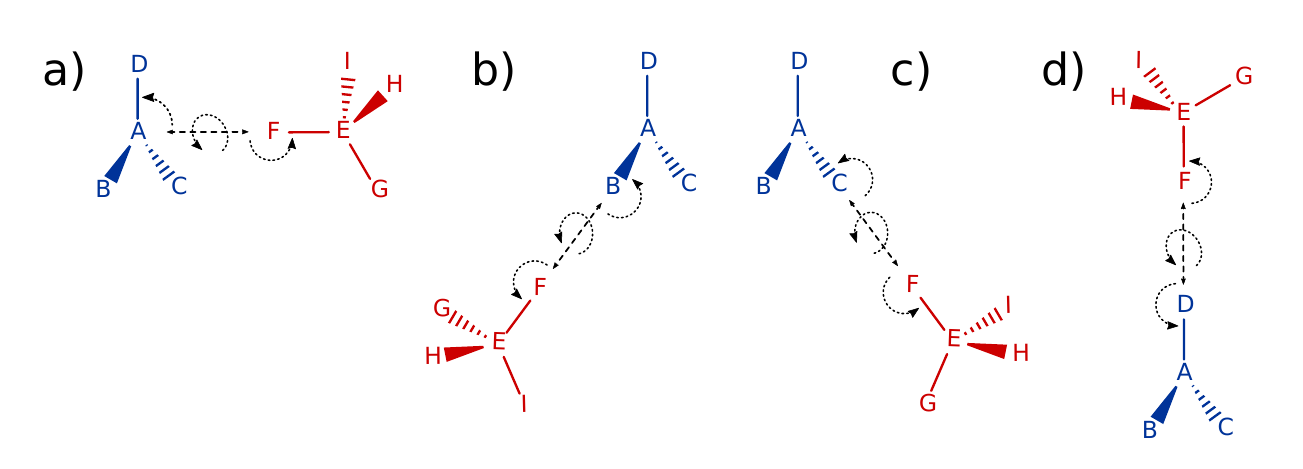}
\caption{
Assembly of reactive complexes between two intermediates colored blue and red.
Three rotational degrees of freedom are indicated by curly arrows.
For clarity, reactive complexes constructed from pairs containing atoms E, G, H, and I are omitted.
}
\label{fig:reactive_complex_exhaustive}
\end{figure}

To exhaustively explore the intermolecular reaction between two intermediates (i.e., all possible products and their reaction paths),
the following steps are carried out.
Firstly, to explore the reaction between any pair of atoms from the intermediates they need to be positioned relative
to one another with the aim of obtaining a \textit{reactive complex} (see Fig.~\ref{fig:reactive_complex_exhaustive}).
Here, their relative orientation (three rotational degrees of freedom) must be considered.
This is particularly important for reactions in which non-covalent bonding is important
or where no single pair of reacting atoms can be defined (as, for example, in a Diels-Alder reaction).
Note also that our restriction to two intermediates does not exclude reactions with a molecularity higher than two as an intermediate
may consist of more than one molecule, which is also important when considering microsolvated structures.

For an exhaustive exploration, reactive complexes must then be generated for every pair of atoms.
However, it is obvious that this is, in general, not feasible (see Section \ref{subsec:reactive_atoms} for a viable solution of this problem).
Finally, through a constrained optimization along a shrinking distance between the reacting atoms, an approximate reaction path is constructed.
A full geometry optimization is carried out afterwards so that either new species are formed or the reactants are recovered,
which is automatically detected.

For an intramolecular reaction, the minimum structure acts as a starting point for the constrained optimization.
The relative orientation of the reacting atoms in the intermediate will determine the reaction path and product.

For both, intra- and intermolecular reactions, the conformational diversity of the intermediates needs
to be taken into account to ensure the exploration of all reaction paths and products.

\subsection{Identification of Reactive Atoms}
\label{subsec:reactive_atoms}

Fig.~\ref{fig:reactive_complex_exhaustive} implies that it is, in general, unfeasible to consider the reaction between
every pair of atoms of two intermediates under full orientational freedom as the complete pairing would lead to a myriad of reactions
most of which potentially featuring high reaction barriers.
Due to the exponential growth of the number of possible reactions highly systematic exploration algorithms such as GRRM\cite{Maeda2013} reach their limits of feasibility.
Therefore, a descriptor is required that allows one to identify pairs of atomic centers that, when brought together in close proximity, are likely to react.
At the same time, the choice of descriptor must not compromise the exhaustiveness of the exploration, that is
it must not confine the exploration to known, expected reaction paths.
Hence, the descriptor should be based on fundamental physical quantities evaluated in a quantum mechanical framework such as the electron density.

When considering the reactivity of spatially extended reactants, descriptors are appropriate that are based on
first principles such as the Laplacian of the electron density,\cite{Bader1994}
Fukui functions,\cite{Fukui1982} partial atomic charges,\cite{Mulliken1955,Mulliken1955a,Meister1994,Bultinck2007} atomic polarizabilities,\cite{Brown1982,Kang1982,Kunz1996}
or dual descriptors\cite{Morell2005,Morell2006,Ayers2007,Cardenas2009}
(see also Refs.~\onlinecite{Geerlings2003,Geerlings2008,Johnson2011} for reviews).
However, all of them suffer from the limitation that it is difficult to assess the height of reaction barriers from information at the reactants' minimum structures,
for which they are evaluated.
Nonetheless, it is a usually very fruitful assumption in chemistry that the minimum structure holds some information on the system's
reactivity,
which is reflected in the considerable success of chemical concepts and of
expert systems applied for synthesis planing.\cite{Corey1976,Pensak1977,Gasteiger1990,Rucker2004,Fialkowski2005,Todd2005,Chen2008,Chen2009,Segler2017,Segler2017a}

\begin{figure}[H]
\centering
\includegraphics{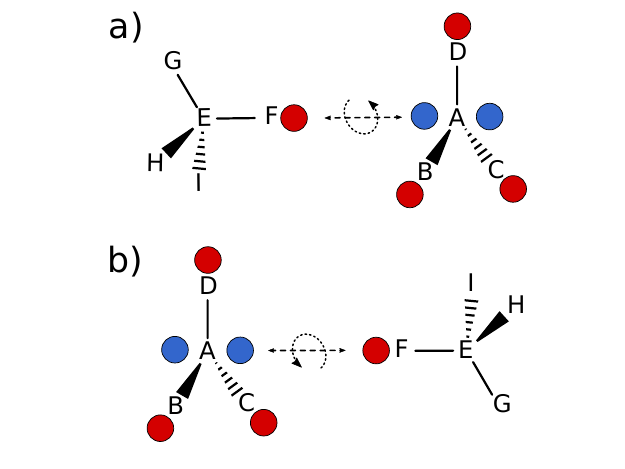}
\caption{
Assembly of reactive complexes after identification of reactive sites.
Atoms A and F are arranged such that reactive sites
of opposite reactivity (discs colored blue and red) are facing each other.
The rotational degrees of freedom of the reactive complexes are indicated by curly arrows.
}
\label{fig:reactive_complex_selection}
\end{figure}

In this work, we pursue an approach that combines basic chemical knowledge and physical principles (such as attraction of oppositely
charged residues) with information extracted from quantum mechanical quantities.
When formulating such heuristic rules one faces a trade-off between efficiency and transferability.
Our descriptors then determine the location of \textit{reactive sites} situated around atoms (depicted as discs in Fig.~\ref{fig:reactive_complex_selection}).
To restrict the exploration to reactions that are likely to feature surmountable reaction barriers under the reaction conditions given,
only pairs of atoms with reactive sites of opposite reactivity are considered.
In the case of an intermolecular reaction, reactants are oriented such that reactive sites are facing each other (see Fig.~\ref{fig:reactive_complex_selection}).
Clearly, these restrictions can be easily lifted in our algorithm to guarantee a successively expanding exploration. This is very much in the spirit of
a rolling exploration of reaction networks that also allows for changing reaction conditions.

\subsection{Exploration of Minimum-Energy Paths and Transition States}
\label{subsec:mep}

From the reactive complex and the reaction product a minimum-energy path connecting them is to be found.
Double-ended transition state search methods such as nudged elastic band\cite{Henkelman2000,Henkelman2000a,Trygubenko2004}
and string methods\cite{E2002,Peters2004,Behn2011,Zimmerman2013,Zimmerman2015a,Jafari2017} are efficient in suggesting an initial guess for a transition state.
A single-ended search method, such as eigenvector following,\cite{Cerjan1981,Simons1983,Wales1992,Wales1993,Jensen1995,Kumeda2001,Bergeler2015a} is then employed to optimize the transition-state candidate
so that a stationary point with exactly one negative eigenvalue of the Hessian matrix is found.
The corresponding eigenvector is followed in the forward and backward directions (by a steepest-descent method) to connect to two local minima.

\begin{figure}[!ht]
\centering
\includegraphics{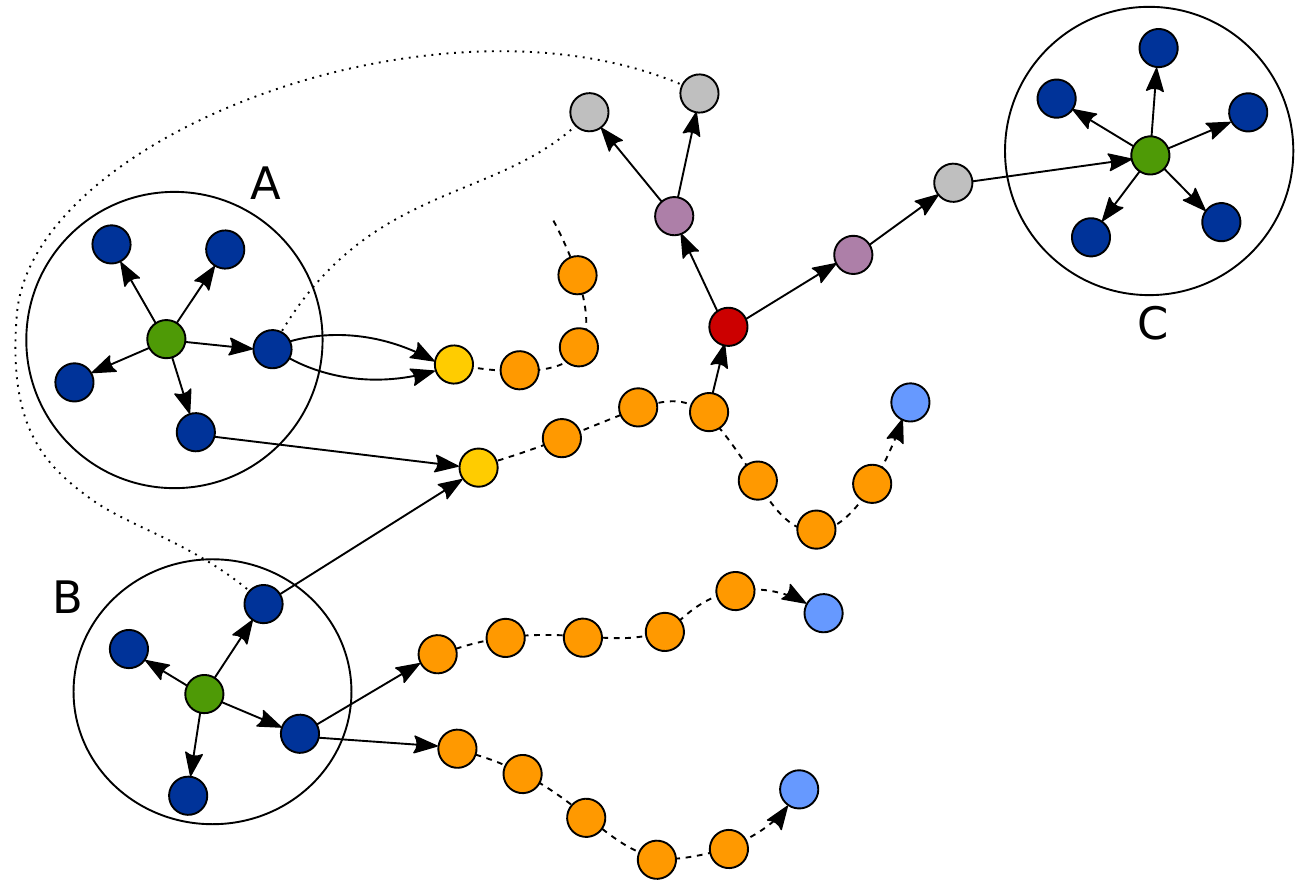}
\caption{
Illustration of the exploration protocol.
Nodes (discs) represent molecular structures.
Conformers of the same configurational isomer (A, B, and C) are enclosed in a circle.
Conformers are generated (dark blue) from an initial conformer (green).
When two conformers react a reactive complexes (yellow) is formed.
For both, inter- and intramolecular reactions, an approximate reaction path (orange nodes, dashed line) is explored.
The last point of the approximate reaction path is optimized to yield a reaction product (light blue).
If the product is different from the reactants, a transition state (red) will be searched for.
An IRC calculation is performed to obtain the two ends of the minimum-energy path (purple).
Minimum energy structures are split into individual molecular structures (gray).
Finally, it is determined whether these gray structures are new (solid arrow)
or whether they are part of the existing network (dotted line).
}
\label{fig:protocol}
\end{figure}

Employing the Mayer bond-order criterion,
minimum-energy structures consisting of more than one molecule are split into separate molecules.
Here, the charge to be assigned to each molecule is determined by calculating the atomic partial charges in the minimum-energy structure.
Finally, through the application of the bond-order criterion it is determined whether the molecules have been encountered before in the exploration
(Fig.~\ref{fig:protocol} illustrates the entire protocol).

This protocol is applied repeatedly until no new structures are explored or the exploration reaches some specified bound
(e.g., determined by a maximum molar mass for an intermediate).
In an advanced set-up, this bound is given by thermodynamic ensemble parameters such as temperature and pressure.
Then, a kinetic simulation would allow one to identify intermediates which are not significantly populated
and to exclude those from subsequent steps of the exploration.

\subsection{Construction of a Reaction Network}
\label{subsec:reaction_network}

From the explored intermediates and transition states a focused network consisting of nodes
(representing molecular configurations) and edges (representing reaction channels) needs to be constructed.
This step is critical for the understanding of the underlying chemical processes
and for carrying out additional analyses such as molecular property calculations and kinetic studies.

\begin{figure}[H]
\centering
\includegraphics{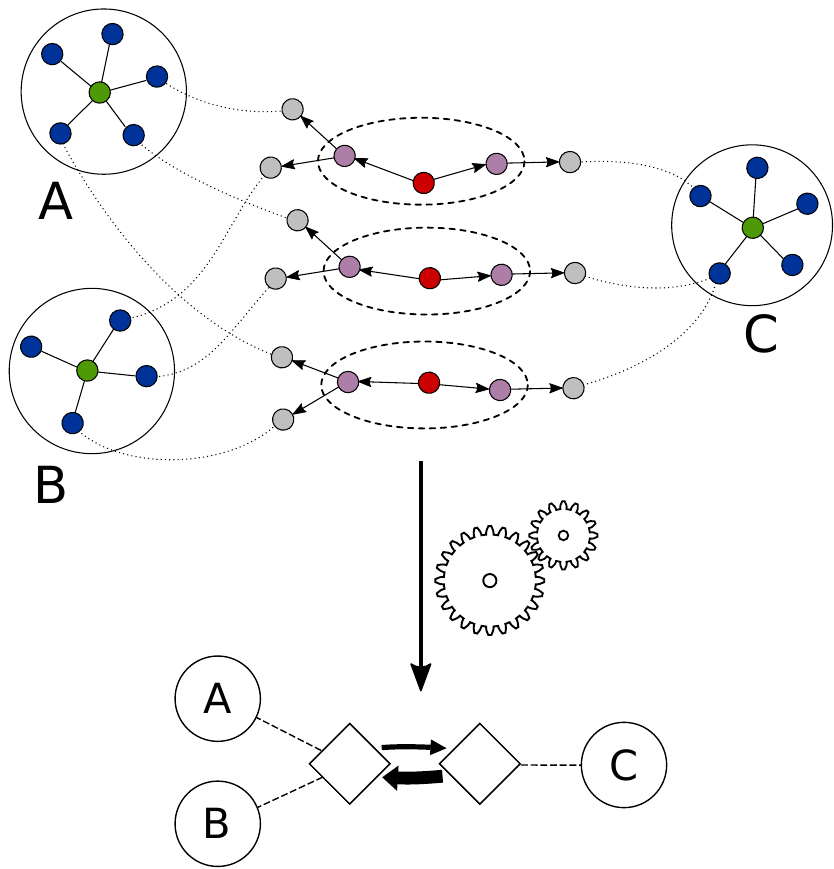}
\caption{
Construction of a reaction network by reduction of the raw exploration network.
\textit{Top}:
molecular configurations (circles) A and B react to form molecular configuration C.
Three minimum-energy paths (dashed ovals) are shown, each consisting of a transition state (red node) and two minimum-energy structures (purple nodes).
The minimum-energy structures are split into their molecular structures (gray nodes) which correspond to conformers (blue nodes)
of the molecular configurations.
\textit{Bottom}:
molecular configurations A and B are placed in a virtual flask (diamond, left), react and form a different virtual flask (diamond, right) consisting of a
molecular configuration C.
}
\label{fig:layout}
\end{figure}

In Fig.~\ref{fig:layout}, the reduction of the raw network to a compact and accessible format is illustrated.
Of the raw network (shown in Fig.~\ref{fig:protocol}) the molecular configurations (labeled A, B, and C) including their conformers (blue nodes),
the transition states (red nodes), the minimum structures of the minimum-energy paths (purple nodes), and the molecular structures they consist of (gray nodes)
are of interest (see Fig.~\ref{fig:layout}, top).
There usually exist multiple reaction paths with different barrier heights for the same chemical transformation (in Fig.~\ref{fig:layout}, $A + B \rightleftarrows C$).
This multitude of reaction paths arises from the consideration of conformational degrees of freedom of the reactants (see Section~\ref{subsec:conformers}) and
the rotational degrees of freedom of the reactive complexes (see Section~\ref{subsec:reactive_complex}).
This situation is shown in Fig.~\ref{fig:layout} (top) by three reaction paths.

In Fig.~\ref{fig:layout} (bottom), a compact representation of the raw network is given.
Molecular configurations A and B are placed in a \textit{virtual flask} (diamond, left in Fig.~\ref{fig:layout}) and
react to form a different virtual flask (diamond, right in Fig.~\ref{fig:layout}) which consists of one molecular configuration C.
The thickness of the arrow between two virtual flasks is proportional to the effective rate constant of the reaction.
The calculation of the effective rate constant from multiple reaction paths is not straightforward and will be
discussed in detail in a forthcoming study
(see also our recent work in Ref.\ \onlinecite{Proppe2016}).
In this study, the thickness of the arrow between two virtual flasks is determined from the height of the lowest activation
barrier of the reaction paths: high barriers are represented by thin arrows, low barriers by thick arrows.

\section{Application to the Formose Reaction}

The formose reaction features complex network structure and product distribution.\cite{Butlerow1861}
The first step in this reaction is the dimerization of formaldehyde to glycolaldehyde which is extremely slow.
It was shown that pure formaldehyde in water is unreactive and that small amounts of contamination are required to initiate the reaction.\cite{Socha1980}
For example, by addition of glycolaldehyde to the reaction mixture the formation of glycolaldehyde and higher sugars is greatly accelerated,
suggesting an autocatalytic mechanism.\cite{Bissette2013}

To explore the formose reaction, we applied our exploration protocol described in Section~\ref{sec:protocol} to an initial state
consisting of formaldehyde, glycolaldehyde, and water.
Since the formose reaction results in an intractable polymeric mixture, we restricted the exploration to a volume in chemical space
that does not exceed the chemical formula of tetrose, i.e., $\text{C}_4 \text{H}_8 \text{O}_4$.

Employing RDKit\cite{RDKit2017033,Riniker2015}, conformers were generated for each molecular configuration (see Section~\ref{subsec:conformers})
according to the protocol described in Ref.~\onlinecite{Ebejer2012}.
To reduce the number of quantum chemical calculations, only the most stable conformer was considered in the subsequent steps of the exploration.

From Mayer bond orders extracted from the electronic wave function\cite{Mayer1983} we constructed a molecular graph
consisting of atoms (vertices) connected by bonds (edges).
Based on arguments of electronegativity, we considered heteroatoms (i.e., oxygen in this system) to be electron-rich.
Hydrogen atoms were considered to have the opposite reactivity if they were found next to a heteroatom within the distance of three edges.
Carbon atoms were considered to feature the reactivity of both, unless they were a neighbor of a heteroatom in which case they were 
automatically labeled electron-poor.
We found that these simple rules work well for the system of organic reactions under consideration here (see below). For future work,
it will be interesting to compare a multitude of descriptors (various concepts evaluated from the electronic wave function such
as partial charges, hardness and softness, electronegativity, the dual descriptor and so forth)
in order to assess their general reliability and transferability
for other types of reaction networks, involving also transition metals.

To form a reactive complex, two reactants were positioned such that atom $i$ of one reactant with a reactive site of $i$ and
atom $j$ of the other reactant with a reactive site of $j$ formed one axis.
Reactive sites of an atom $i$ were located on a sphere centered on $i$ with radius equal to the van der Waals radius of $i$
by maximizing the distance to all neighboring atoms.
Two additional reactive complexes were generated by rotating one reactant around this axis by 120\textdegree{} and 240\textdegree{}
(see Fig.~\ref{fig:reactive_complex_selection}). In principle, the value and number of angles as a means for orientational screening 
can, however, be chosen freely.

Two molecular configurations were compared by finding the maximum common subgraph (MCS) of their graph representations
and stereochemical information was considered.
The MCS was determined by \texttt{RDKit}.\cite{RDKit2017033}

The exploration comprised
82990 geometry optimizations,
23690 constrained PES scans,
7657 freezing-string,
13675 eigenvector following,
and 10458 IRC calculations.
Details on the computational methodology are provided in the appendix.
Note that the transition state guess obtained from the freezing-string calculation may contain more than one imaginary frequency.
As a result, the number of eigenvector following calculations is larger than the number of freezing-string calculations.
In total, 934 unique molecular configurations were identified and 6871 minimum-energy paths connecting them were explored.
The reaction network comprising all structures is given in the Supporting Information.

\subsection{Reaction Network}

\begin{figure}[H]
\centering
\includegraphics[width=\textwidth]{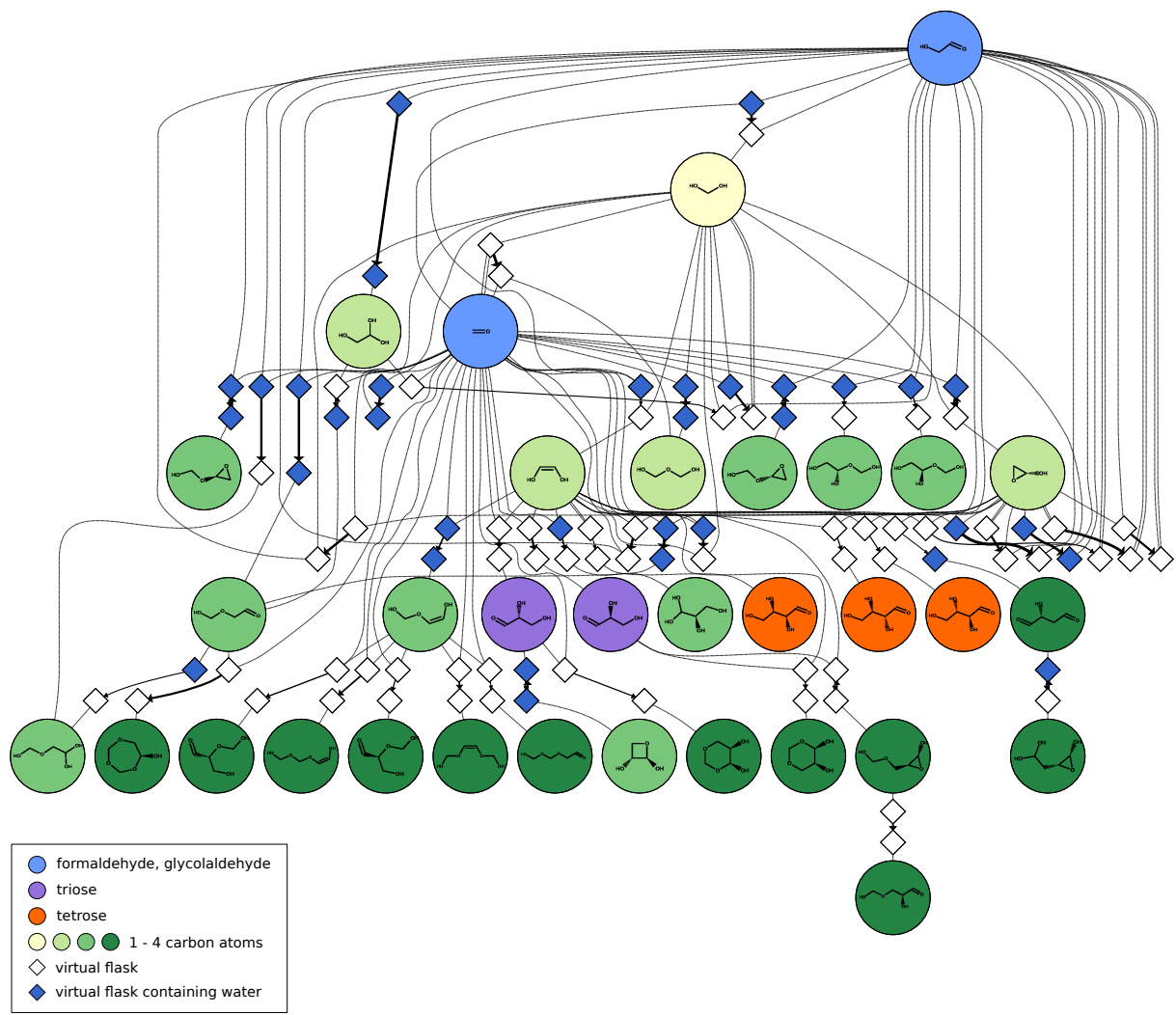}
\caption{
Reaction network generated from formaldehyde, glycolaldehyde, and water (the last not shown explicitly)
consisting of reactions with activation barriers below 50~kJ/mol.
}
\label{fig:network_detailed}
\end{figure}

As described in Section~\ref{subsec:reaction_network}, the raw exploration network was processed to generate a reaction network.
In Fig.~\ref{fig:network_detailed}, the resulting network is shown.
In this network, reactions with barriers above 50~kJ/mol are omitted.
The nodes representing the starting materials formaldehyde and glycolaldehyde are colored light blue.
Water is not shown explicitly, but virtual flasks (diamonds) containing at least one water molecule are colored dark blue.
If not stated otherwise the fill color of disc shaped nodes indicates the number of carbon atoms in the corresponding molecule.

It can be seen that starting from formaldehyde, glycolaldehyde, and water two trioses (green nodes) and three tetrose (orange nodes) can be formed
through multiple cascades of reactions.
In addition, multiple three- to six-membered rings can be identified and even a seven-membered ring is among the products.
It can also be seen that most of the polymerization reactions are irreversible which is in accord with experimental findings.\cite{Butlerow1861,Breslow1959}

Furthermore, one can observe that for some intermediates (e.g., (2R)-oxiran-2-ol)
the corresponding enantiomer is missing in the network.
However, given achiral starting materials, the product should be racemic.
This bias can be explained by the selection of only one conformer for each molecular configuration which was considered
in intra- and intermolecular reactions.
This issue can be easily resolved by considering sufficiently many conformers for each molecular configuration.

\begin{figure}[H]
\centering
\includegraphics[width=\textwidth]{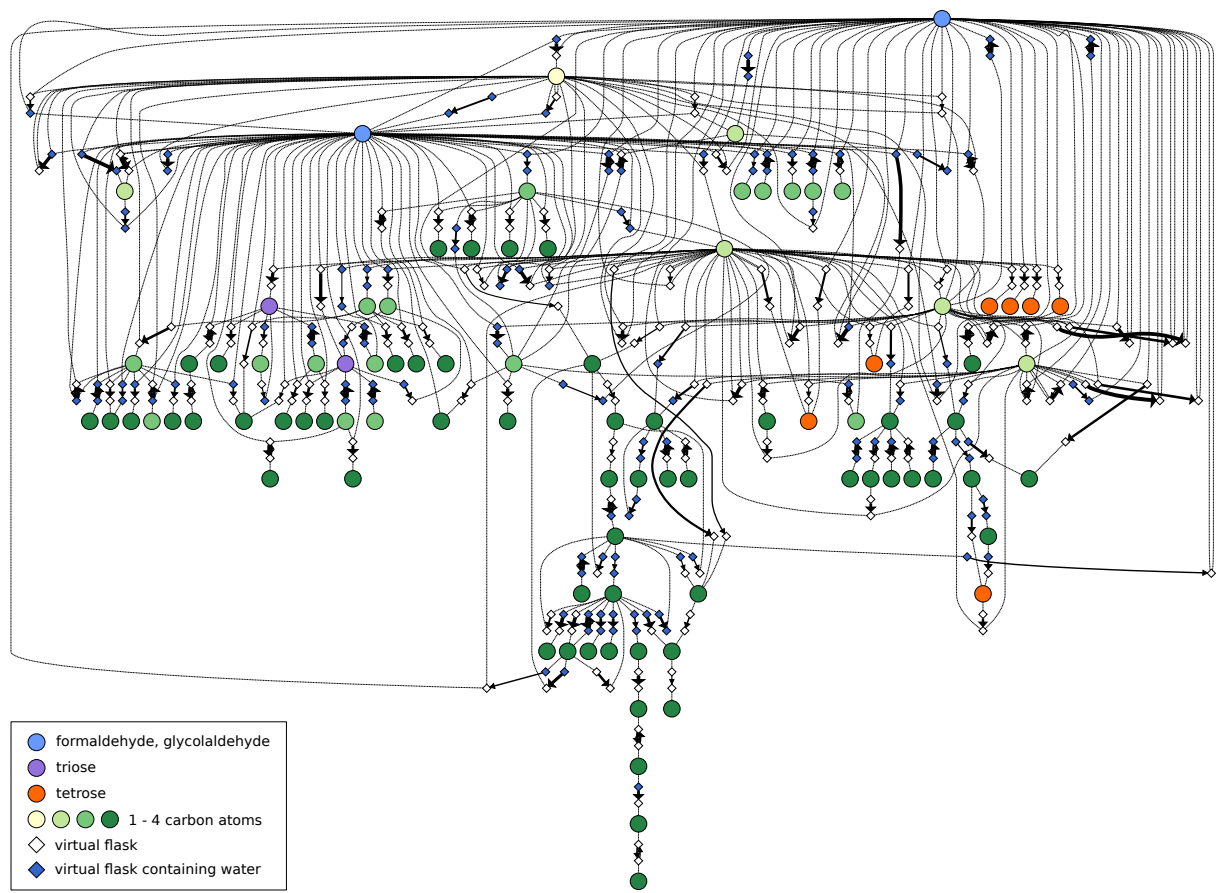}
\caption{
Reaction network generated from formaldehyde, glycolaldehyde, and water (the last not shown explicitly)
consisting of reactions with activation barriers below 85~kJ/mol.
}
\label{fig:network_purist}
\end{figure}

In Fig.~\ref{fig:network_purist}, the reaction network was further expanded to reactions with barriers between 50 and 85~kJ/mol
(reactions networks with a higher cutoff are too large and not suitable for illustration).
It can be clearly seen that the reaction network becomes increasingly complex when considering higher activation barriers.
A reaction network with barriers up to 100~kJ/mol is given in the Supporting Information.

It can also be seen that there are multiple different pathways to form a species and it is not obvious from the barrier heights
which of the pathways are the most dominant ones.
Kinetic simulations and additional data analysis are therefore necessary to reach a deeper understanding of the underlying
reaction process.

Finally, to assess the extensiveness of our protocol we compared our reaction network with one obtained from a manual exploration.\cite{Kua2013}
Within the bounds preset for the present exploration, each intermediate and reaction path identified in a limited manual exploration by
Kua et al.\ \cite{Kua2013} can be found in our reaction network.

\subsection{Alternative Reaction Paths}

Through the consideration of conformational diversity and orientational degrees of freedom in the assembly of reactive complexes,
our exploration protocol aims to explore all potential reaction paths between two intermediates.
The multitude of reaction paths is discovered through the assembly of multiple reactive complexes
(as sketched in Fig.~\ref{fig:reactive_complex_selection}).
The following example shall demonstrate the importance of a thorough exploration of reaction paths.

\begin{figure}[H]
\centering
\includegraphics{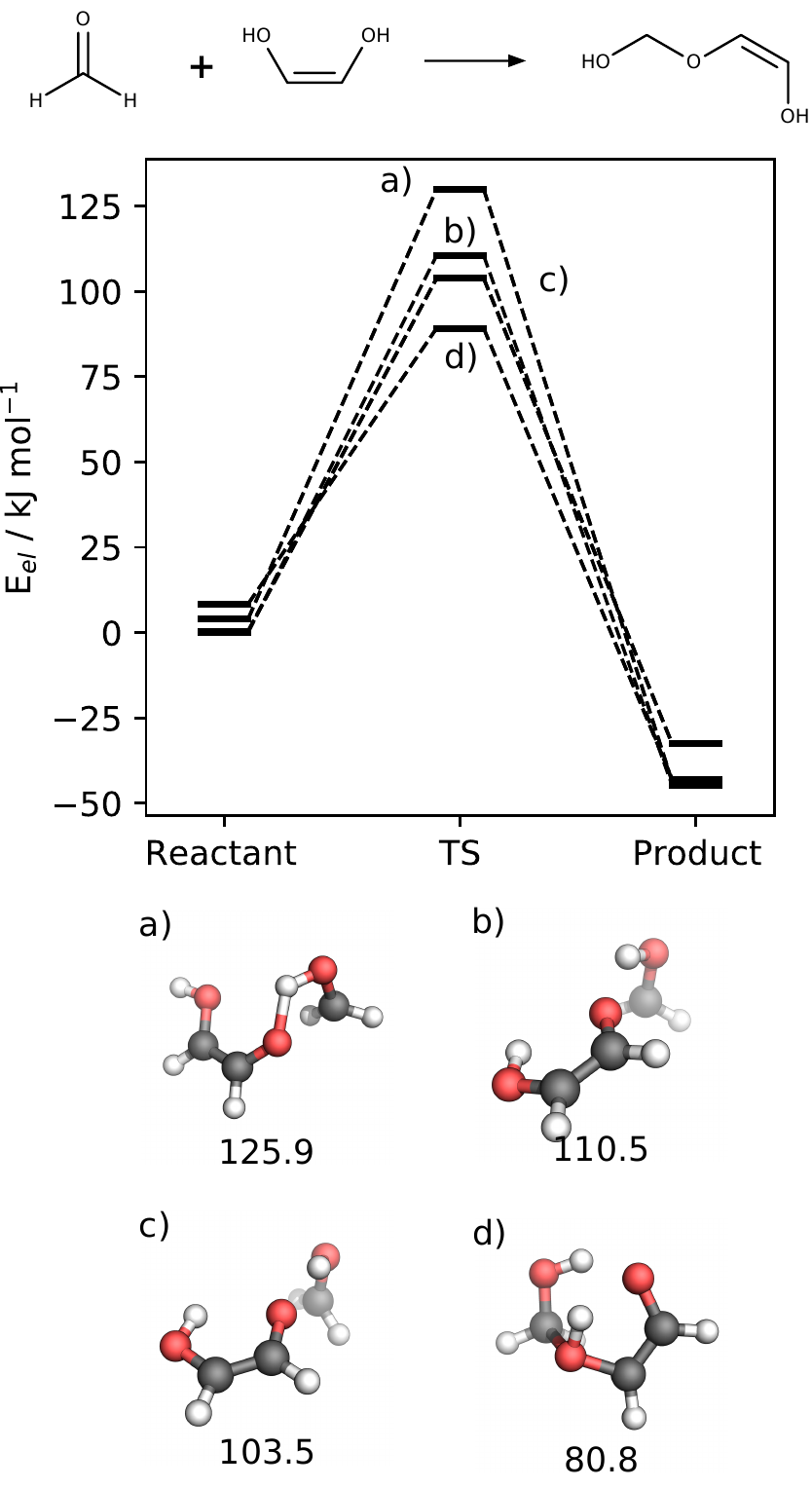}
\caption{
\textit{Top}: reaction profiles of four minimum-energy paths for the reaction between ethene-1,2-diol and formaldehyde
forming 2-(hydroxymethoxy)ethen-1-ol.
\textit{Bottom}: molecular structures of the transition states.
}
\label{fig:barriers_nowater}
\end{figure}

In Fig.~\ref{fig:barriers_nowater}, a selection of minimum-energy paths (from the raw exploration network) for the reaction between ethene-1,2-diol and
formaldehyde forming 2-(hydroxymethoxy)ethen-1-ol is shown.
The barrier heights of the paths in the forward direction range from 80.8 to 125.9~kJ/mol (neglecting solvation effects).
In conventional transition state theory\cite{Eyring1935}, a difference of $\approx 45$ kJ/mol in
the barrier height results in a difference in the reaction rate that is on the order of $10^{8}$ at room temperature
(assuming that the difference in electronic energy solely determines the free energy difference).
Therefore, for kinetic analyses the exploration of all reaction paths is crucial.
It can also be seen that despite the small number of atoms involved in this reaction,
the structures of the transition states differ significantly.
For example, in Fig.~\ref{fig:barriers_nowater} a), the linear arrangement prevents the stabilizing interactions present
in the cyclic transition state shown in Fig.~\ref{fig:barriers_nowater} d).
It is the explicit consideration of rotational degrees of freedom when constructing the reactive complexes that leads
to the uncovering of these reaction paths.

\begin{figure}[H]
\centering
\includegraphics{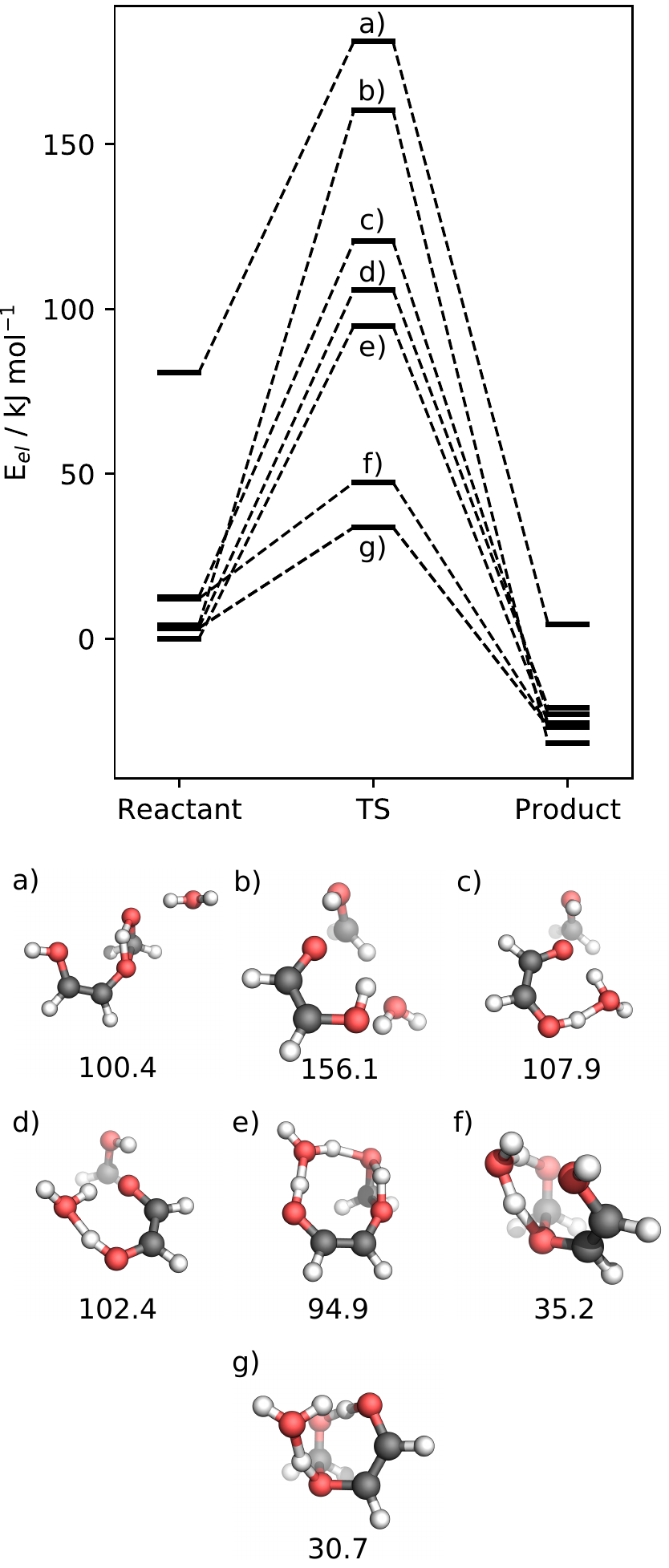}
\caption{
\textit{Top}: reaction profiles of seven minimum-energy paths for the reaction between ethene-1,2-diol and formaldehyde
forming 2-(hydroxymethoxy)ethen-1-ol catalyzed by a water molecule.
\textit{Bottom}: molecular structures of the transition states.
}
\label{fig:barriers_water}
\end{figure}

In the following, the effect of microsolvation on the exploration of reaction paths is investigated.
In Fig.~\ref{fig:barriers_water}, seven paths from the exploration network are shown.
The chemical transformation is the same as in the previous example, but this time in the presence of one water molecule.

It can be seen that, compared to Fig.~\ref{fig:barriers_nowater}, both the number of different reaction paths and
the spread of barrier heights (ranging from 30.7 to 156.1~kJ/mol) increased.
Moreover, Fig.~\ref{fig:barriers_nowater}, e) shows that our exploration protocol is clearly capable of finding reaction 
paths that involve more than two reacting atoms, although all reactive complexes started from the pairing-of-atoms concept.
The plethora of possible transition paths due to the added degrees of freedom of the solvent molecules renders explorations very challenging.
While the application of a continuum model may suffice for unreactive, apolar solvents such as hexane,\cite{Tomasi2005}
for polar solvents exhibiting directional bonding
such as water which may actively participate in the reaction through hydrogen bonding and transfer (as can be seen in Fig.~\ref{fig:barriers_water}, f)
this is not a viable solution.
A hybrid approach in which microsolvated solutes are embedded into a continuum model is a convenient compromise
as long as explicit sampling by molecular dynamics or Monte Carlo methods can be avoided for certain parts of the network.

\subsection{Graph Analysis of Reaction Network}

\begin{figure}[H]
\hspace*{-1cm}\includegraphics[width=1.2\textwidth]{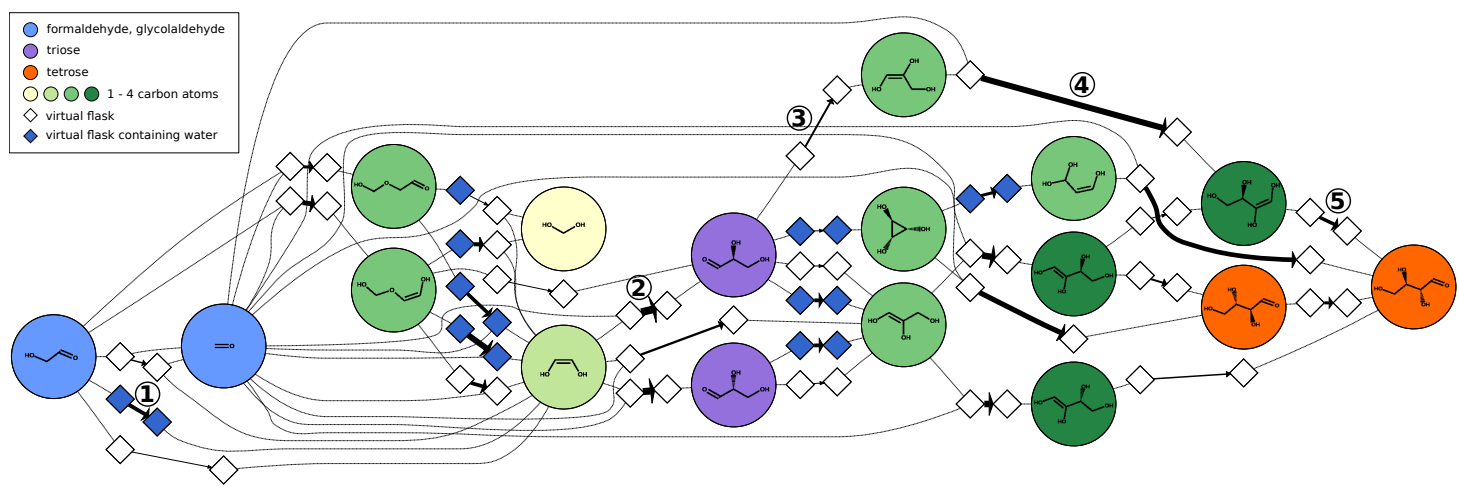}
\caption{
Reaction pathways starting from glycolaldehyde (far left node) leading to the formation of \textsc{d}-erythrose (far right node).
}
\label{fig:erythrose_paths}
\end{figure}

To study the process of sugar formation in the formose reaction we applied a tree traversal algorithm to the reaction network
to find all paths that start from glycolaldehyde and lead to the naturally occurring tetrose \textsc{d}-erythrose.
To take into account the low concentration of all products at the beginning of the reaction,
we took paths only into consideration if in all elementary reactions there was not more than one reactant that was not a starting material.
In addition, edges representing reactions with barriers above 250~kJ/mol were removed.

We were able to identify 40 distinct paths comprising up to five elementary reactions.
Fig.~\ref{fig:erythrose_paths} shows a subnetwork in which each molecular configuration and reaction is present in at least one of these paths.
The elementary reactions of the path with the lowest barrier heights are indicated by the numbers 1 to 5.
With an activation barrier of 190~kJ/mol, the third reaction of this path features the highest barrier.

\begin{figure}[H]
\centering
\includegraphics[width=\textwidth]{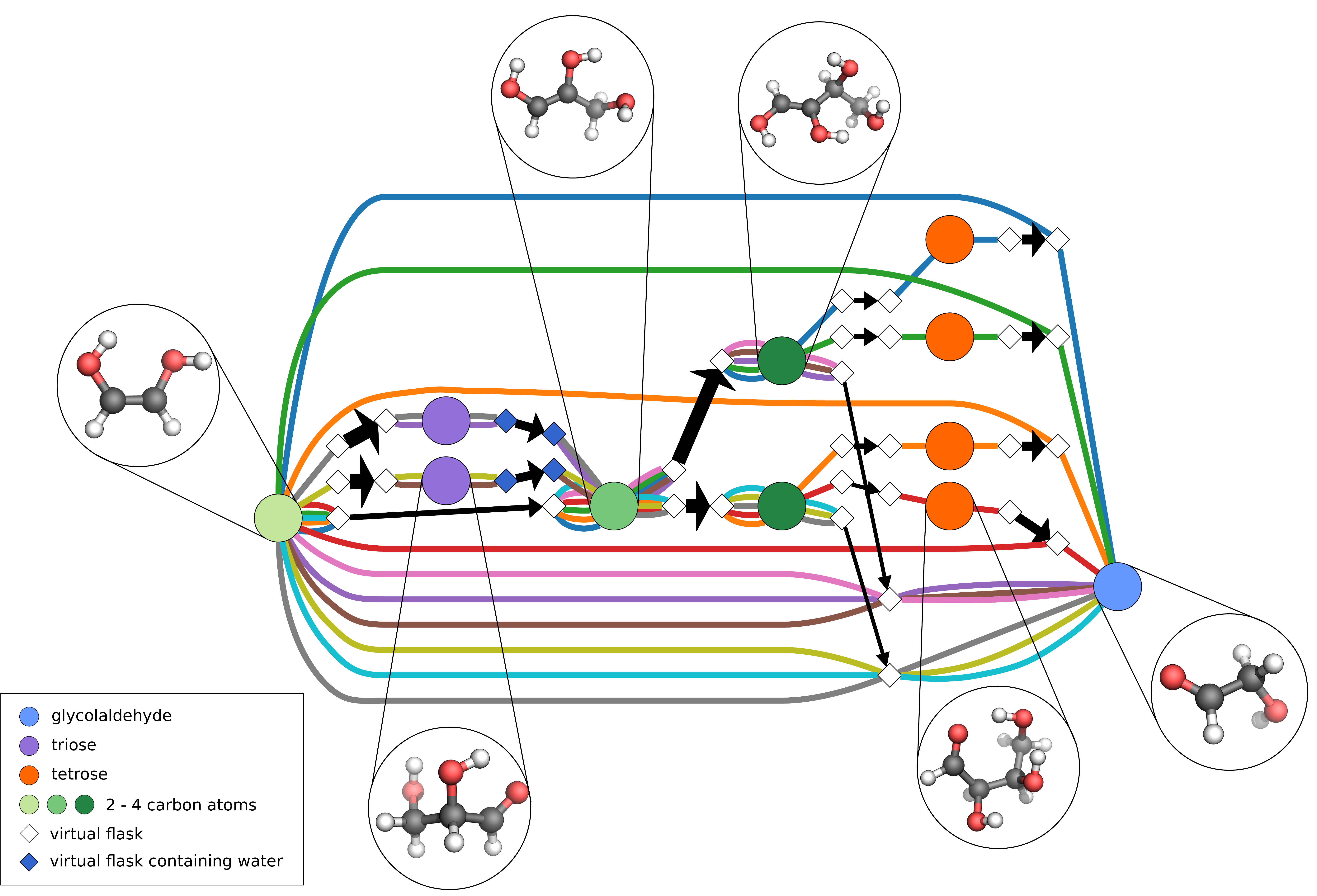}
\caption{
Autocatalytic cycles starting from ethene-1,2-diol (far left node) leading to the formation of glycolaldehyde (far right node).
For clarity, formaldehyde is not shown.
}
\label{fig:cycles}
\end{figure}

Employing tree traversal algorithms we also searched for autocatalytic processes in the reaction network.
Fig.~\ref{fig:cycles} shows a subnetwork consisting of ten cycles (colored solid lines) consisting of up to four elementary reactions
with the lowest reaction barriers starting from ethene-1,2-diol
that lead to the formation of glycolaldehyde through the consecutive addition of formaldehyde.
Ethene-1,2-diol can readily be formed from the starting material via an enolization reaction (see Fig.~\ref{fig:network_detailed}).
For clarity, formaldehyde is not shown in this network.

\begin{figure}[H]
\centering
\includegraphics{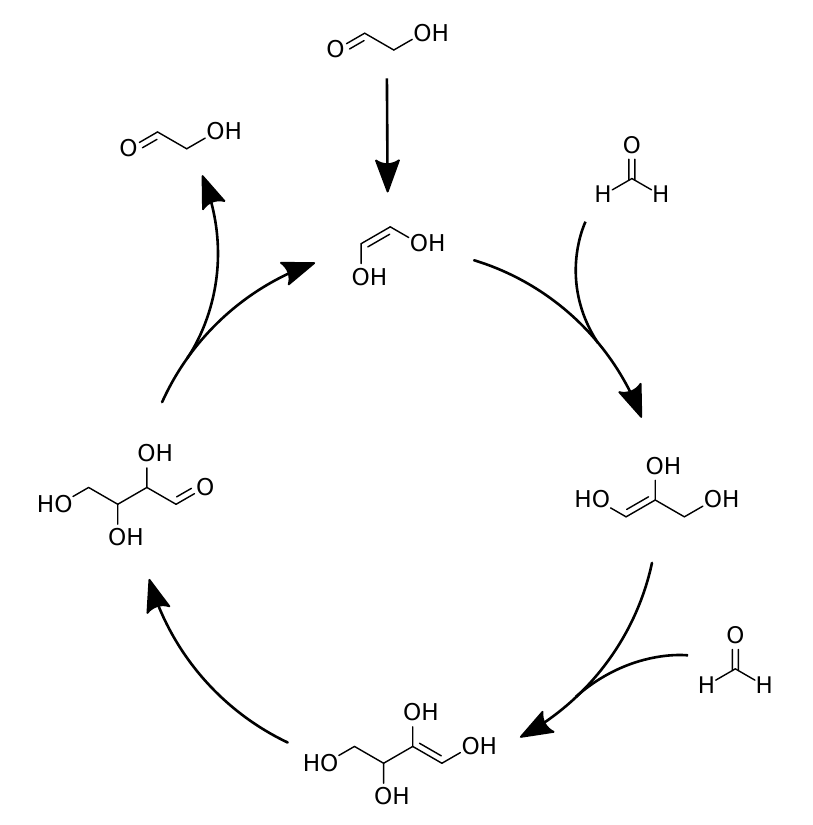}
\caption{
Explored autocatalytic cycle with the lowest activation barriers.
}
\label{fig:catalytic_cycle}
\end{figure}

The cycle with the lowest reaction barriers (dark blue path) is depicted in Fig.~\ref{fig:catalytic_cycle} and involves the formation of prop-1-ene-1,2,3-triol from ethene-1,2-diol and formaldehyde,
followed by the formation of a compound consisting of four carbons (dark green node).
A subsequent enolization reaction yields a tetrose which undergoes a fragmentation reaction in which glycolaldehyde is produced and ethene-1,2-diol is recovered.
It can also be seen that both trioses and all four tetroses are formed in multiple autocatalytic cycles.

\section{Conclusions}

In this work, a new robust and generally applicable protocol for the fully automated exploration of complex chemical reactions is presented.
By addressing limitations of our previous approach\cite{Bergeler2015} we were able to explore a chemical system
in unprecedented depth.
Starting from a set of initial conditions (i.e., starting material, solvent, and temperature),
an exploration network is built and extended through the repeated application of our protocol.
New intermediates are explored through the construction of reactive complexes between already explored ones, new reactants, and intramolecular reactions.
By applying heuristic rules based on conceptual electronic-structure theory, bond orders, and graph theoretical considerations,
we were able to tame the combinatorial explosion of possible reactive complexes.
Subsequently, a reaction was induced in each reactive complex and an approximate reaction path was explored.
To ensure a thorough exploration of these paths, we considered rotational degrees of freedom during the construction of reactive complexes.
Through transition state optimizations, the approximate paths were refined to minimum-energy paths.
The exploration network was then processed to afford a compressed reaction network consisting of molecular configurations
and reaction channels connecting them.
The reaction network was then visualized by an automatically generated graph structure.

We applied our protocol to the formose reaction, a prebiotic oligomerization reaction resulting in a plethora of products, including monosaccharides.
We explored a vast number of intermediates and minimum-energy paths to yield a reaction network featuring complex network structure and product distribution.
Through the application of tree traversal algorithms, different pathways leading to the formation of naturally occurring tetrose \textsc{d}-erythrose were identified.
Furthermore, we discovered multiple pathways in the reaction network that rationalize the autocatalytic properties of the formose reaction.
In addition, we showed that there can exist many minimum-energy paths with different reaction barriers for the same chemical transformation.
Many of these reaction paths would remain undiscovered in a tedious manual exploration attempt.

For an even further improved description of the formose reaction the following aspects need to be addressed.
Firstly, an appropriate solvent model and the calculation of thermodynamic properties is mandatory.
Secondly, the full conformational complexity needs to be taken into account.
For some intermediates with many floppy degrees of freedom (e.g., polymers) \textit{ab initio} molecular dynamics simulations or Monte Carlo
configurational sampling
might be more suitable for a sufficiently rigorous exploration of local minima and may therefore be employed in specific sections of the growing reaction network.
Furthermore, bond dissociation reactions and accompanying kinetic analyses need to be incorporated in the rolling exploration protocol.
Work along these lines is in progress in our laboratory.

\section*{Appendix: Computational Methodology}
\label{sec:comp_methodology}

The exploration was carried out by our new program package \texttt{Chemoton} in a fully automated fashion.
The exploration protocol was implemented in C++.
We plan to make \texttt{Chemoton} publicly available in the near future through our web pages.

All quantum chemical calculations were performed with the \texttt{Q-Chem} program package (version 4.3)\cite{Shao2015} employing
the PBE exchange-correlation functional\cite{Perdew1996a} and a double-$\zeta$ basis.\cite{Dunning1970}
We emphasize that our exploration protocol works with any electronic structure method and we chose a density-functional model
for the sake of convenience. Also, for the raw-data generation other quantum chemistry packages can be easily interfaced.

For single point calculations, structure optimizations, and vibrational analyses default settings were kept.
The maximum number of self-consistent field calculations was set to 1200 for structure optimizations, potential energy scans,
freezing-string calculations, transition state searches, and IRC calculations.
For potential energy scans the convergence on energy change of successive optimization cycles was set to $10^{-4}$ Hartree.
In freezing-string calculations the number of nodes was chosen to be 20 and the number of perpendicular gradient steps was six.
For transition state searches the maximum allowed step size was reduced to 0.05~$\r{A}$.

Activation barriers were approximated by the difference in electronic energy of reactant and transition state, i.e.,
vibrational corrections were not included (this may be easily changed in standard approximations valid for the gas phase,
but will increase the computational effort for the raw data generation).
It is also easy to switch on a continuum solvation model in the exploration. However, we were interested in the exploration
of a generic network first and therefore did not switch on dielectric continuum solvation. Solvation can then be studied in a subsequent
step, where the network is copied multiple times to account for different solvation environments (also considering extensive
microsolvation and configuration-space sampling as an intermediary layer between the solute and the continuum embedding)
into which each isolated-structure node is then automatically embedded. Further refinement of these networks within
the solvation model may be considered afterwards.

All calculations and the progress of the exploration was saved to a Mongo database.\cite{MongoDB32}
Automated data analysis was performed with the Python libraries \texttt{matplotlib}\cite{Hunter2007} and \texttt{pandas}.\cite{McKinney2010}
The graphical representation of the reaction network were created by the \texttt{Graphviz} program.\cite{Gansner2000}

\section*{Acknowledgments}

This work has been financially supported by the Schweizerischer Nationalfonds (Project No. 20020\_169120).
GNS gratefully acknowledges support by a PhD fellowship of the Fonds der Chemischen Industrie.

\section*{Supporting Information}

The complete exploration data containing all Cartesian coordinates and an extensive version of the graphical representation of the network
can be found in the supporting information.
This information is available free of charge via the Internet at http://pubs.acs.org/.


\begin{thebibliography}{100}

\bibitem{Masters2011}
Masters,~C. \textit{Homogeneous {{Transition}}-Metal {{Catalysis}}: {{A Gentle
  Art}};} {Springer}: 1981 edition ed.; 2011.

\bibitem{Vinu2012}
Vinu,~R.;\ \ Broadbelt,~L.~J.  Unraveling {{Reaction Pathways}} and
  {{Specifying Reaction Kinetics}} for {{Complex Systems}},  \textit{Annu. Rev.
  Chem. Biomol. Eng.} \textbf{2012,} \textsl{3,} 29--54.

\bibitem{Vereecken2015}
Vereecken,~L.;\ \ Glowacki,~D.~R.;\ \ Pilling,~M.~J.  Theoretical {{Chemical
  Kinetics}} in {{Tropospheric Chemistry}}: {{Methodologies}} and
  {{Applications}},  \textit{Chem. Rev.} \textbf{2015,} \textsl{115,}
  4063--4114.

\bibitem{Saitta2014}
Saitta,~A.~M.;\ \ Saija,~F.  Miller Experiments in Atomistic Computer
  Simulations,  \textit{Proc. Natl. Acad. Sci.} \textbf{2014,} \textsl{111,}
  13768--13773.

\bibitem{Wang2014}
Wang,~L.-P.;\ \ Titov,~A.;\ \ McGibbon,~R.;\ \ Liu,~F.;\ \ Pande,~V.~S.;\ \
  Mart{\'\i}nez,~T.~J.  Discovering Chemistry with an Ab Initio Nanoreactor,
  \textit{Nat. Chem.} \textbf{2014,} \textsl{6,} 1044--1048.

\bibitem{Wang2016}
Wang,~L.-P.;\ \ McGibbon,~R.~T.;\ \ Pande,~V.~S.;\ \ Martinez,~T.~J.  Automated
  {{Discovery}} and {{Refinement}} of {{Reactive Molecular Dynamics Pathways}},
   \textit{J. Chem. Theory Comput.} \textbf{2016,} \textsl{12,} 638--649.

\bibitem{vanDuin2001}
{van Duin},~A. C.~T.;\ \ Dasgupta,~S.;\ \ Lorant,~F.;\ \ Goddard,~W.~A.
  {{ReaxFF}}:\, {{A Reactive Force Field}} for {{Hydrocarbons}},  \textit{J.
  Phys. Chem. A} \textbf{2001,} \textsl{105,} 9396--9409.

\bibitem{Dontgen2015}
D{\"o}ntgen,~M.;\ \ Przybylski-Freund,~M.-D.;\ \ Kr{\"o}ger,~L.~C.;\ \
  Kopp,~W.~A.;\ \ Ismail,~A.~E.;\ \ Leonhard,~K.  Automated {{Discovery}} of
  {{Reaction Pathways}}, {{Rate Constants}}, and {{Transition States Using
  Reactive Molecular Dynamics Simulations}},  \textit{J. Chem. Theory Comput.}
  \textbf{2015,} \textsl{11,} 2517--2524.

\bibitem{Fischer1992}
Fischer,~S.;\ \ Karplus,~M.  Conjugate Peak Refinement: An Algorithm for
  Finding Reaction Paths and Accurate Transition States in Systems with Many
  Degrees of Freedom,  \textit{Chem. Phys. Lett.} \textbf{1992,} \textsl{194,}
  252--261.

\bibitem{Florian2003}
Flori{\'a}n,~J.;\ \ Goodman,~M.~F.;\ \ Warshel,~A.  Computer {{Simulation}} of
  the {{Chemical Catalysis}} of {{DNA Polymerases}}:\, {{Discriminating}}
  between {{Alternative Nucleotide Insertion Mechanisms}} for {{T7 DNA
  Polymerase}},  \textit{J. Am. Chem. Soc.} \textbf{2003,} \textsl{125,}
  8163--8177.

\bibitem{Garcia-Viloca2004}
Garcia-Viloca,~M.;\ \ Gao,~J.;\ \ Karplus,~M.;\ \ Truhlar,~D.~G.  How {{Enzymes
  Work}}: {{Analysis}} by {{Modern Rate Theory}} and {{Computer Simulations}},
  \textit{Science} \textbf{2004,} \textsl{303,} 186--195.

\bibitem{Imhof2009}
Imhof,~P.;\ \ Fischer,~S.;\ \ Smith,~J.~C.  Catalytic {{Mechanism}} of {{DNA
  Backbone Cleavage}} by the {{Restriction Enzyme EcoRV}}: {{A Quantum
  Mechanical}}/{{Molecular Mechanical Analysis}},  \textit{Biochemistry}
  \textbf{2009,} \textsl{48,} 9061--9075.

\bibitem{Reidelbach2016}
Reidelbach,~M.;\ \ Betz,~F.;\ \ M{\"a}usle,~R.~M.;\ \ Imhof,~P.  Proton
  Transfer Pathways in an Aspartate-Water Cluster Sampled by a Network of
  Discrete States,  \textit{Chem. Phys. Lett.} \textbf{2016,} \textsl{659,}
  169--175.

\bibitem{Imhof2016}
Imhof,~P.  A {{Networks Approach}} to {{Modeling Enzymatic Reactions}},
  \textit{Methods Enzymol.} \textbf{2016,} \textsl{578,} 249--271.

\bibitem{Senn2006}
Senn,~H.~M.;\ \ Thiel,~W.  {{QM}}/{{MM Methods}} for {{Biological Systems}}.
  In  \textit{Atomistic {{Approaches}} in {{Modern Biology}}}; Topics in
  Current Chemistry {Springer, Berlin, Heidelberg}: 2006.

\bibitem{Senn2007}
Senn,~H.~M.;\ \ Thiel,~W.  {{QM}}/{{MM}} Studies of Enzymes,  \textit{Curr.
  Opin. Chem. Biol.} \textbf{2007,} \textsl{11,} 182--187.

\bibitem{Senn2009}
Senn,~H.~M.;\ \ Thiel,~W.  {{QM}}/{{MM Methods}} for {{Biomolecular Systems}},
  \textit{Angew. Chem. Int. Ed.} \textbf{2009,} \textsl{48,} 1198--1229.

\bibitem{Ohno2004}
Ohno,~K.;\ \ Maeda,~S.  A Scaled Hypersphere Search Method for the Topography
  of Reaction Pathways on the Potential Energy Surface,  \textit{Chem. Phys.
  Lett.} \textbf{2004,} \textsl{384,} 277--282.

\bibitem{Ohno2008}
Ohno,~K.;\ \ Maeda,~S.  Automated Exploration of Reaction Channels,
  \textit{Phys. Scr.} \textbf{2008,} \textsl{78,} 058122.

\bibitem{Maeda2013}
Maeda,~S.;\ \ Ohno,~K.;\ \ Morokuma,~K.  Systematic Exploration of the
  Mechanism of Chemical Reactions: The Global Reaction Route Mapping ({{GRRM}})
  Strategy Using the {{ADDF}} and {{AFIR}} Methods,  \textit{Phys. Chem. Chem.
  Phys.} \textbf{2013,} \textsl{15,} 3683--3701.

\bibitem{Satoh2015}
Satoh,~H.;\ \ Oda,~T.;\ \ Nakakoji,~K.;\ \ Uno,~T.;\ \ Iwata,~S.;\ \ Ohno,~K.
  "{{Maizo}}"-Chemistry {{Project}}: Toward {{Molecular}}- and {{Reaction
  Discovery}} from {{Quantum Mechanical Global Reaction Route Mappings}},
  \textit{J. Comput. Chem. Jpn.} \textbf{2015,} \textsl{14,} 77--79.

\bibitem{Satoh2016}
Satoh,~H.;\ \ Oda,~T.;\ \ Nakakoji,~K.;\ \ Uno,~T.;\ \ Tanaka,~H.;\ \
  Iwata,~S.;\ \ Ohno,~K.  Potential {{Energy Surface}}-{{Based Automatic
  Deduction}} of {{Conformational Transition Networks}} and {{Its Application}}
  on {{Quantum Mechanical Landscapes}} of d-{{Glucose Conformers}},  \textit{J.
  Chem. Theory Comput.} \textbf{2016,} \textsl{12,} 5293--5308.

\bibitem{Ohno2017}
Ohno,~K.;\ \ Kishimoto,~N.;\ \ Iwamoto,~T.;\ \ Satoh,~H.  Global Exploration of
  Isomers and Isomerization Channels on the Quantum Chemical Potential Energy
  Surface of {{H3CNO3}},  \textit{J. Comput. Chem.} \textbf{2017,} \textsl{38,}
  669--687.

\bibitem{Levy2017}
Levy,~D.~E. \textit{Arrow-{{Pushing}} in {{Organic Chemistry}}: {{An Easy
  Approach}} to {{Understanding Reaction Mechanisms}};} {Wiley}: 2nd edition
  ed.; 2017.

\bibitem{Rucker2004}
R{\"u}cker,~C.;\ \ R{\"u}cker,~G.;\ \ Bertz,~S.~H.  Organic {{Synthesis}} -
  {{Art}} or {{Science}}?,  \textit{J. Chem. Inf. Comput. Sci.} \textbf{2004,}
  \textsl{44,} 378--386.

\bibitem{Todd2005}
Todd,~M.~H.  Computer-Aided Organic Synthesis,  \textit{Chem. Soc. Rev.}
  \textbf{2005,} \textsl{34,} 247--266.

\bibitem{Chen2008}
Chen,~J.~H.;\ \ Baldi,~P.  Synthesis {{Explorer}}: {{A Chemical Reaction
  Tutorial System}} for {{Organic Synthesis Design}} and {{Mechanism
  Prediction}},  \textit{J. Chem. Educ.} \textbf{2008,} \textsl{85,} 1699.

\bibitem{Chen2009}
Chen,~J.~H.;\ \ Baldi,~P.  No {{Electron Left Behind}}: {{A Rule}}-{{Based
  Expert System To Predict Chemical Reactions}} and {{Reaction Mechanisms}},
  \textit{J. Chem. Inf. Model.} \textbf{2009,} \textsl{49,} 2034--2043.

\bibitem{Graulich2010}
Graulich,~N.;\ \ Hopf,~H.;\ \ Schreiner,~P.~R.  Heuristic Thinking Makes a
  Chemist Smart,  \textit{Chem. Soc. Rev.} \textbf{2010,} \textsl{39,}
  1503--1512.

\bibitem{Kayala2011}
Kayala,~M.~A.;\ \ Azencott,~C.-A.;\ \ Chen,~J.~H.;\ \ Baldi,~P.  Learning to
  {{Predict Chemical Reactions}},  \textit{J. Chem. Inf. Model.} \textbf{2011,}
  \textsl{51,} 2209--2222.

\bibitem{Kayala2012}
Kayala,~M.~A.;\ \ Baldi,~P.  {{ReactionPredictor}}: {{Prediction}} of {{Complex
  Chemical Reactions}} at the {{Mechanistic Level Using Machine Learning}},
  \textit{J. Chem. Inf. Model.} \textbf{2012,} \textsl{52,} 2526--2540.

\bibitem{Zimmerman2013a}
Zimmerman,~P.~M.  Automated Discovery of Chemically Reasonable Elementary
  Reaction Steps,  \textit{J. Comput. Chem.} \textbf{2013,} \textsl{34,}
  1385--1392.

\bibitem{Rappoport2014}
Rappoport,~D.;\ \ Galvin,~C.~J.;\ \ Zubarev,~D.~Y.;\ \ Aspuru-Guzik,~A.
  Complex {{Chemical Reaction Networks}} from {{Heuristics}}-{{Aided Quantum
  Chemistry}},  \textit{J. Chem. Theory Comput.} \textbf{2014,} \textsl{10,}
  897--907.

\bibitem{Zimmerman2015}
Zimmerman,~P.~M.  Navigating Molecular Space for Reaction Mechanisms: An
  Efficient, Automated Procedure,  \textit{Mol. Simul.} \textbf{2015,}
  \textsl{41,} 43--54.

\bibitem{Zubarev2015}
Zubarev,~D.~Y.;\ \ Rappoport,~D.;\ \ Aspuru-Guzik,~A.  Uncertainty of
  {{Prebiotic Scenarios}}: {{The Case}} of the {{Non}}-{{Enzymatic Reverse
  Tricarboxylic Acid Cycle}},  \textit{Sci. Rep.} \textbf{2015,} \textsl{5,}
  1--7.

\bibitem{Habershon2015}
Habershon,~S.  Sampling Reactive Pathways with Random Walks in Chemical Space:
  {{Applications}} to Molecular Dissociation and Catalysis,  \textit{J. Chem.
  Phys.} \textbf{2015,} \textsl{143,} 094106.

\bibitem{Habershon2016}
Habershon,~S.  Automated {{Prediction}} of {{Catalytic Mechanism}} and {{Rate
  Law Using Graph}}-{{Based Reaction Path Sampling}},  \textit{J. Chem. Theory
  Comput.} \textbf{2016,} \textsl{12,} 1786--1798.

\bibitem{Pendleton2016}
Pendleton,~I.~M.;\ \ P{\'e}rez-Temprano,~M.~H.;\ \ Sanford,~M.~S.;\ \
  Zimmerman,~P.~M.  Experimental and {{Computational Assessment}} of
  {{Reactivity}} and {{Mechanism}} in {{C}}(Sp3)\textendash{}{{N
  Bond}}-{{Forming Reductive Elimination}} from {{Palladium}}({{IV}}),
  \textit{J. Am. Chem. Soc.} \textbf{2016,} \textsl{138,} 6049--6060.

\bibitem{Ludwig2016}
Ludwig,~J.~R.;\ \ Zimmerman,~P.~M.;\ \ Gianino,~J.~B.;\ \ Schindler,~C.~S.
  Iron({{III}})-Catalysed Carbonyl\textendash{}olefin Metathesis,
  \textit{Nature} \textbf{2016,} \textsl{533,} 374--379.

\bibitem{Dewyer2017}
Dewyer,~A.~L.;\ \ Zimmerman,~P.~M.  Finding Reaction Mechanisms, Intuitive or
  Otherwise,  \textit{Chem. Soc. Rev.} \textbf{2017,} \textsl{15,} 501--504.

\bibitem{Butlerow1861}
Butlerow,~A.  Bildung Einer Zuckerartigen {{Substanz}} Durch {{Synthese}},
  \textit{Justus Liebigs Ann. Chem.} \textbf{1861,} \textsl{120,} 295--298.

\bibitem{Bergeler2015}
Bergeler,~M.;\ \ Simm,~G.~N.;\ \ Proppe,~J.;\ \ Reiher,~M.  Heuristics-{{Guided
  Exploration}} of {{Reaction Mechanisms}},  \textit{J. Chem. Theory Comput.}
  \textbf{2015,} \textsl{11,} 5712--5722.

\bibitem{Becke1990}
Becke,~A.~D.;\ \ Edgecombe,~K.~E.  A Simple Measure of Electron Localization in
  Atomic and Molecular Systems,  \textit{J. Chem. Phys.} \textbf{1990,}
  \textsl{92,} 5397--5403.

\bibitem{Fukui1970}
Fukui,~K.  Formulation of the Reaction Coordinate,  \textit{J. Phys. Chem.}
  \textbf{1970,} \textsl{74,} 4161--4163.

\bibitem{Yandulov2003}
Yandulov,~D.~V.;\ \ Schrock,~R.~R.;\ \ Rheingold,~A.~L.;\ \ Ceccarelli,~C.;\ \
  Davis,~W.~M.  Synthesis and {{Reactions}} of {{Molybdenum Triamidoamine
  Complexes Containing Hexaisopropylterphenyl Substituents}},  \textit{Inorg.
  Chem.} \textbf{2003,} \textsl{42,} 796--813.

\bibitem{Yandulov2003a}
Yandulov,~D.~V.;\ \ Schrock,~R.~R.  Catalytic {{Reduction}} of {{Dinitrogen}}
  to {{Ammonia}} at a {{Single Molybdenum Center}},  \textit{Science}
  \textbf{2003,} \textsl{301,} 76--78.

\bibitem{Ganti1975}
G{\'a}nti,~T.  Organization of Chemical Reactions into Dividing and
  Metabolizing Units: {{The}} Chemotons,  \textit{Biosystems} \textbf{1975,}
  \textsl{7,} 15--21.

\bibitem{Eschenmoser1992}
Eschenmoser,~A.;\ \ Loewenthal,~E.  Chemistry of Potentially Prebiological
  Natural Products,  \textit{Chem. Soc. Rev.} \textbf{1992,} \textsl{21,}
  1--16.

\bibitem{Delidovich2014}
Delidovich,~I.~V.;\ \ Simonov,~A.~N.;\ \ Taran,~O.~P.;\ \ Parmon,~V.~N.
  Catalytic {{Formation}} of {{Monosaccharides}}: {{From}} the {{Formose
  Reaction}} towards {{Selective Synthesis}},  \textit{ChemSusChem}
  \textbf{2014,} \textsl{7,} 1833--1846.

\bibitem{Decker1982}
Decker,~P.;\ \ Schweer,~H.;\ \ Pohlamnn,~R.  Bioids,  \textit{J. Chromatogr.,
  A} \textbf{1982,} \textsl{244,} 281--291.

\bibitem{Zweckmair2014}
Zweckmair,~T.;\ \ B{\"o}hmdorfer,~S.;\ \ Bogolitsyna,~A.;\ \ Rosenau,~T.;\ \
  Potthast,~A.;\ \ Novalin,~S.  Accurate {{Analysis}} of {{Formose Reaction
  Products}} by {{LC}}\textendash{}{{UV}}: {{An Analytical Challenge}},
  \textit{J. Chromatogr. Sci.} \textbf{2014,} \textsl{52,} 169--175.

\bibitem{Breslow1959}
Breslow,~R.  On the Mechanism of the Formose Reaction,  \textit{Tetrahedron
  Lett.} \textbf{1959,} \textsl{1,} 22--26.

\bibitem{Bissette2013}
Bissette,~A.~J.;\ \ Fletcher,~S.~P.  Mechanisms of {{Autocatalysis}},
  \textit{Angew. Chem. Int. Ed.} \textbf{2013,} \textsl{52,} 12800--12826.

\bibitem{Kim2011}
Kim,~H.-J.;\ \ Ricardo,~A.;\ \ Illangkoon,~H.~I.;\ \ Kim,~M.~J.;\ \
  Carrigan,~M.~A.;\ \ Frye,~F.;\ \ Benner,~S.~A.  Synthesis of
  {{Carbohydrates}} in {{Mineral}}-{{Guided Prebiotic Cycles}},  \textit{J. Am.
  Chem. Soc.} \textbf{2011,} \textsl{133,} 9457--9468.

\bibitem{Proppe2016}
Proppe,~J.;\ \ Husch,~T.;\ \ Simm,~G.~N.;\ \ Reiher,~M.  Uncertainty
  Quantification for Quantum Chemical Models of Complex Reaction Networks,
  \textit{Faraday Discuss.} \textbf{2016,} \textsl{195,} 497--520.

\bibitem{Vainio2007}
Vainio,~M.~J.;\ \ Johnson,~M.~S.  Generating {{Conformer Ensembles Using}} a
  {{Multiobjective Genetic Algorithm}},  \textit{J. Chem. Inf. Model.}
  \textbf{2007,} \textsl{47,} 2462--2474.

\bibitem{Leite2007}
Leite,~T.~B.;\ \ Gomes,~D.;\ \ Miteva,~M.~A.;\ \ Chomilier,~J.;\ \
  Villoutreix,~B.~O.;\ \ Tuff{\'e}ry,~P.  Frog: A {{FRee Online druG 3D}}
  Conformation Generator,  \textit{Nucleic Acids Res.} \textbf{2007,}
  \textsl{35,} W568--W572.

\bibitem{O'Boyle2011}
O'Boyle,~N.~M.;\ \ Vandermeersch,~T.;\ \ Flynn,~C.~J.;\ \ Maguire,~A.~R.;\ \
  Hutchison,~G.~R.  Confab - {{Systematic}} Generation of Diverse Low-Energy
  Conformers,  \textit{J. Cheminf.} \textbf{2011,} \textsl{3,} 8.

\bibitem{Miteva2010}
Miteva,~M.~A.;\ \ Guyon,~F.;\ \ Tuff{\'e}ry,~P.  Frog2: {{Efficient 3D}}
  Conformation Ensemble Generator for Small Compounds,  \textit{Nucleic Acids
  Res.} \textbf{2010,} \textsl{38,} W622--W627.

\bibitem{Riniker2015}
Riniker,~S.;\ \ Landrum,~G.~A.  Better {{Informed Distance Geometry}}: {{Using
  What We Know To Improve Conformation Generation}},  \textit{J. Chem. Inf.
  Model.} \textbf{2015,} \textsl{55,} 2562--2574.

\bibitem{Mayer1983}
Mayer,~I.  Charge, Bond Order and Valence in the {{AB}} Initio {{SCF}} Theory,
  \textit{Chem. Phys. Lett.} \textbf{1983,} \textsl{97,} 270--274.

\bibitem{Bader1994}
Bader,~R. F.~W. \textit{Atoms in {{Molecules}}: {{A Quantum Theory}};}
  {Clarendon Press}: Oxford England : New York, 1994.

\bibitem{Fukui1982}
Fukui,~K.  Role of {{Frontier Orbitals}} in {{Chemical Reactions}},
  \textit{Science} \textbf{1982,} \textsl{218,} 747--754.

\bibitem{Mulliken1955}
Mulliken,~R.~S.  Electronic {{Population Analysis}} on
  {{LCAO}}\textendash{}{{MO Molecular Wave Functions}}. {{I}},  \textit{J.
  Chem. Phys.} \textbf{1955,} \textsl{23,} 1833--1840.

\bibitem{Mulliken1955a}
Mulliken,~R.~S.  Electronic {{Population Analysis}} on
  {{LCAO}}\textendash{}{{MO Molecular Wave Functions}}. {{II}}. {{Overlap
  Populations}}, {{Bond Orders}}, and {{Covalent Bond Energies}},  \textit{J.
  Chem. Phys.} \textbf{1955,} \textsl{23,} 1841--1846.

\bibitem{Meister1994}
Meister,~J.;\ \ Schwarz,~W. H.~E.  Principal {{Components}} of {{Ionicity}},
  \textit{J. Phys. Chem.} \textbf{1994,} \textsl{98,} 8245--8252.

\bibitem{Bultinck2007}
Bultinck,~P.;\ \ Van~Alsenoy,~C.;\ \ Ayers,~P.~W.;\ \ Carb{\'o}-Dorca,~R.
  Critical Analysis and Extension of the {{Hirshfeld}} Atoms in Molecules,
  \textit{J. Chem. Phys.} \textbf{2007,} \textsl{126,} 144111.

\bibitem{Brown1982}
Brown,~R.~E.;\ \ Simas,~A.~M.  On the Applicability of {{CNDO}} Indices for the
  Prediction of Chemical Reactivity,  \textit{Theoret. Chim. Acta}
  \textbf{1982,} \textsl{62,} 1--16.

\bibitem{Kang1982}
Kang,~Y.~K.;\ \ Jhon,~M.~S.  Additivity of Atomic Static Polarizabilities and
  Dispersion Coefficients,  \textit{Theoret. Chim. Acta} \textbf{1982,}
  \textsl{61,} 41--48.

\bibitem{Kunz1996}
Kunz,~C.~F.;\ \ H{\"a}ttig,~C.;\ \ Hess,~B.~A.  Ab Initio Study of the
  Individual Interaction Energy Components in the Ground State of the Mercury
  Dimer,  \textit{Mol. Phys.} \textbf{1996,} \textsl{89,} 139--156.

\bibitem{Morell2005}
Morell,~C.;\ \ Grand,~A.;\ \ Toro-Labb{\'e},~A.  New {{Dual Descriptor}} for
  {{Chemical Reactivity}},  \textit{J. Phys. Chem. A} \textbf{2005,}
  \textsl{109,} 205--212.

\bibitem{Morell2006}
Morell,~C.;\ \ Grand,~A.;\ \ Toro-Labb{\'e},~A.  Theoretical Support for Using
  the {{$\Delta$f}}(r) Descriptor,  \textit{Chem. Phys. Lett.} \textbf{2006,}
  \textsl{425,} 342--346.

\bibitem{Ayers2007}
Ayers,~P.;\ \ Morell,~C.;\ \ De\hspace{0.25em}Proft,~F.;\ \ Geerlings,~P.
  Understanding the {{Woodward}}\textendash{}{{Hoffmann Rules}} by {{Using
  Changes}} in {{Electron Density}},  \textit{Chem. Eur. J.} \textbf{2007,}
  \textsl{13,} 8240--8247.

\bibitem{Cardenas2009}
C{\'a}rdenas,~C.;\ \ Rabi,~N.;\ \ Ayers,~P.~W.;\ \ Morell,~C.;\ \
  Jaramillo,~P.;\ \ Fuentealba,~P.  Chemical {{Reactivity Descriptors}} for
  {{Ambiphilic Reagents}}: {{Dual Descriptor}}, {{Local Hypersoftness}}, and
  {{Electrostatic Potential}},  \textit{J. Phys. Chem. A} \textbf{2009,}
  \textsl{113,} 8660--8667.

\bibitem{Geerlings2003}
Geerlings,~P.;\ \ De~Proft,~F.;\ \ Langenaeker,~W.  Conceptual {{Density
  Functional Theory}},  \textit{Chem. Rev.} \textbf{2003,} \textsl{103,}
  1793--1874.

\bibitem{Geerlings2008}
Geerlings,~P.;\ \ Proft,~F.~D.  Conceptual {{DFT}}: The Chemical Relevance of
  Higher Response Functions,  \textit{Phys. Chem. Chem. Phys.} \textbf{2008,}
  \textsl{10,} 3028--3042.

\bibitem{Johnson2011}
Johnson,~P.~A.;\ \ Bartolotti,~L.~J.;\ \ Ayers,~P.~W.;\ \ Fievez,~T.;\ \
  Geerlings,~P.  Charge {{Density}} and {{Chemical Reactions}}: {{A Unified
  View}} from {{Conceptual DFT}}.   In  \textit{Modern {{Charge}}-{{Density
  Analysis}}}; {Springer, Dordrecht}: 2011.

\bibitem{Corey1976}
Corey,~E.~J.;\ \ Jorgensen,~W.~L.  Computer-Assisted Synthetic Analysis.
  {{Synthetic}} Strategies Based on Appendages and the Use of Reconnective
  Transforms,  \textit{J. Am. Chem. Soc.} \textbf{1976,} \textsl{98,} 189--203.

\bibitem{Pensak1977}
Pensak,~D.~A.;\ \ Corey,~E.~J.  {{LHASA}} - {{Logic}} and {{Heuristics
  Applied}} to {{Synthetic Analysis}}.   In  \textit{Computer-{{Assisted
  Organic Synthesis}}}, Vol.~61; {American Chemical Society}: 1977.

\bibitem{Gasteiger1990}
Gasteiger,~J.;\ \ Ihlenfeldt,~W.~D.  The {{WODCA System}}.   In
  \textit{Software {{Development}} in {{Chemistry}} 4}; Gasteiger,~P. D.~J.,\ \
  Ed.;  {Springer Berlin Heidelberg}: 1990.

\bibitem{Fialkowski2005}
Fialkowski,~M.;\ \ Bishop,~K. J.~M.;\ \ Chubukov,~V.~A.;\ \ Campbell,~C.~J.;\ \
  Grzybowski,~B.~A.  Architecture and {{Evolution}} of {{Organic Chemistry}},
  \textit{Angew. Chem. Int. Ed.} \textbf{2005,} \textsl{44,} 7263--7269.

\bibitem{Segler2017}
Segler,~M. H.~S.;\ \ Waller,~M.~P.  Neural-{{Symbolic Machine Learning}} for
  {{Retrosynthesis}} and {{Reaction Prediction}},  \textit{Chem. Eur. J.}
  \textbf{2017,} \textsl{23,} 5966--5971.

\bibitem{Segler2017a}
Segler,~M. H.~S.;\ \ Waller,~M.~P.  Modelling {{Chemical Reasoning}} to
  {{Predict}} and {{Invent Reactions}},  \textit{Chem. Eur. J.} \textbf{2017,}
  \textsl{23,} 6118--6128.

\bibitem{Henkelman2000}
Henkelman,~G.;\ \ Uberuaga,~B.~P.;\ \ J{\'o}nsson,~H.  A Climbing Image Nudged
  Elastic Band Method for Finding Saddle Points and Minimum Energy Paths,
  \textit{J. Chem. Phys.} \textbf{2000,} \textsl{113,} 9901--9904.

\bibitem{Henkelman2000a}
Henkelman,~G.;\ \ J{\'o}nsson,~H.  Improved Tangent Estimate in the Nudged
  Elastic Band Method for Finding Minimum Energy Paths and Saddle Points,
  \textit{J. Chem. Phys.} \textbf{2000,} \textsl{113,} 9978--9985.

\bibitem{Trygubenko2004}
Trygubenko,~S.~A.;\ \ Wales,~D.~J.  A Doubly Nudged Elastic Band Method for
  Finding Transition States,  \textit{J. Chem. Phys.} \textbf{2004,}
  \textsl{120,} 2082--2094.

\bibitem{E2002}
E,~W.;\ \ Ren,~W.;\ \ Vanden-Eijnden,~E.  String Method for the Study of Rare
  Events,  \textit{Phys. Rev. B} \textbf{2002,} \textsl{66,} 052301.

\bibitem{Peters2004}
Peters,~B.;\ \ Heyden,~A.;\ \ Bell,~A.~T.;\ \ Chakraborty,~A.  A Growing String
  Method for Determining Transition States: {{Comparison}} to the Nudged
  Elastic Band and String Methods,  \textit{J. Chem. Phys.} \textbf{2004,}
  \textsl{120,} 7877--7886.

\bibitem{Behn2011}
Behn,~A.;\ \ Zimmerman,~P.~M.;\ \ Bell,~A.~T.;\ \ Head-Gordon,~M.  Efficient
  Exploration of Reaction Paths via a Freezing String Method,  \textit{J. Chem.
  Phys.} \textbf{2011,} \textsl{135,} 224108.

\bibitem{Zimmerman2013}
Zimmerman,~P.  Reliable {{Transition State Searches Integrated}} with the
  {{Growing String Method}},  \textit{J. Chem. Theory Comput.} \textbf{2013,}
  \textsl{9,} 3043--3050.

\bibitem{Zimmerman2015a}
Zimmerman,~P.~M.  Single-Ended Transition State Finding with the Growing String
  Method,  \textit{J. Comput. Chem.} \textbf{2015,} \textsl{36,} 601--611.

\bibitem{Jafari2017}
Jafari,~M.;\ \ Zimmerman,~P.~M.  Reliable and Efficient Reaction Path and
  Transition State Finding for Surface Reactions with the Growing String
  Method,  \textit{J. Comput. Chem.} \textbf{2017,} \textsl{38,} 645--658.

\bibitem{Cerjan1981}
Cerjan,~C.~J.;\ \ Miller,~W.~H.  On Finding Transition States,  \textit{J.
  Chem. Phys.} \textbf{1981,} \textsl{75,} 2800--2806.

\bibitem{Simons1983}
Simons,~J.;\ \ Joergensen,~P.;\ \ Taylor,~H.;\ \ Ozment,~J.  Walking on
  Potential Energy Surfaces,  \textit{J. Phys. Chem.} \textbf{1983,}
  \textsl{87,} 2745--2753.

\bibitem{Wales1992}
Wales,~D.~J.  Basins of Attraction for Stationary Points on a Potential-Energy
  Surface,  \textit{J. Chem. Soc., Faraday Trans.} \textbf{1992,} \textsl{88,}
  653--657.

\bibitem{Wales1993}
Wales,~D.~J.  Locating Stationary Points for Clusters in Cartesian Coordinates,
   \textit{J. Chem. Soc., Faraday Trans.} \textbf{1993,} \textsl{89,}
  1305--1313.

\bibitem{Jensen1995}
Jensen,~F.  Locating Transition Structures by Mode Following: {{A}} Comparison
  of Six Methods on the {{Ar8 Lennard}}-{{Jones}} Potential,  \textit{J. Chem.
  Phys.} \textbf{1995,} \textsl{102,} 6706--6718.

\bibitem{Kumeda2001}
Kumeda,~Y.;\ \ Wales,~D.~J.;\ \ Munro,~L.~J.  Transition States and
  Rearrangement Mechanisms from Hybrid Eigenvector-Following and Density
  Functional Theory.,  \textit{Chem. Phys. Lett.} \textbf{2001,} \textsl{341,}
  185--194.

\bibitem{Bergeler2015a}
Bergeler,~M.;\ \ Herrmann,~C.;\ \ Reiher,~M.  Mode-Tracking Based
  Stationary-Point Optimization,  \textit{J. Comput. Chem.} \textbf{2015,}
  \textsl{36,} 1429--1438.

\bibitem{Socha1980}
Socha,~R.~F.;\ \ Weiss,~A.~H.;\ \ Sakharov,~M.~M.  Autocatalysis in the Formose
  Reaction,  \textit{React. Kinet. Catal. Lett.} \textbf{1980,} \textsl{14,}
  119--128.

\bibitem{RDKit2017033}
{Gregory Landrum}, ``{{RDKit}} 2017.03.3'',  \url{http://www.rdkit.org/},
  visited on 2017-07-09.

\bibitem{Ebejer2012}
Ebejer,~J.-P.;\ \ Morris,~G.~M.;\ \ Deane,~C.~M.  Freely {{Available Conformer
  Generation Methods}}: {{How Good Are They}}?,  \textit{J. Chem. Inf. Model.}
  \textbf{2012,} \textsl{52,} 1146--1158.

\bibitem{Kua2013}
Kua,~J.;\ \ Avila,~J.~E.;\ \ Lee,~C.~G.;\ \ Smith,~W.~D.  Mapping the
  {{Kinetic}} and {{Thermodynamic Landscape}} of {{Formaldehyde
  Oligomerization}} under {{Neutral Conditions}},  \textit{J. Phys. Chem. A}
  \textbf{2013,} \textsl{117,} 12658--12667.

\bibitem{Eyring1935}
Eyring,~H.  The {{Activated Complex}} in {{Chemical Reactions}},  \textit{J.
  Chem. Phys.} \textbf{1935,} \textsl{3,} 107--115.

\bibitem{Tomasi2005}
Tomasi,~J.;\ \ Mennucci,~B.;\ \ Cammi,~R.  Quantum {{Mechanical Continuum
  Solvation Models}},  \textit{Chem. Rev.} \textbf{2005,} \textsl{105,}
  2999--3094.

\bibitem{Shao2015}
Shao,~Y. \textit{et al.}\   Advances in Molecular Quantum Chemistry Contained
  in the {{Q}}-{{Chem}} 4 Program Package,  \textit{Mol. Phys.} \textbf{2015,}
  \textsl{113,} 184--215.

\bibitem{Perdew1996a}
Perdew,~J.~P.;\ \ Burke,~K.;\ \ Ernzerhof,~M.  Generalized {{Gradient
  Approximation Made Simple}},  \textit{Phys. Rev. Lett.} \textbf{1996,}
  \textsl{77,} 3865--3868.

\bibitem{Dunning1970}
Dunning,~T.~H.  Gaussian {{Basis Functions}} for {{Use}} in {{Molecular
  Calculations}}. {{I}}. {{Contraction}} of (9s5p) {{Atomic Basis Sets}} for
  the {{First}}-{{Row Atoms}},  \textit{J. Chem. Phys.} \textbf{1970,}
  \textsl{53,} 2823--2833.

\bibitem{MongoDB32}
{MongoDB Inc.}, ``{{MongoDB}} 3.2'',  \url{www.mongodb.com}, visited on
  2017-07-01.

\bibitem{Hunter2007}
Hunter,~J.~D.  Matplotlib: {{A 2D}} Graphics Environment,  \textit{Comput. Sci.
  Eng.} \textbf{2007,} \textsl{9,} 90--95.

\bibitem{McKinney2010}
McKinney,~W.  Data {{Structures}} for {{Statistical Computing}} in {{Python}}.
   In  \textit{Proceedings of the 9th {{Python}} in {{Science Conference}}};
  van~der Walt,~S.;\ \ Millman,~J.,\ \ Eds.;  2010.

\bibitem{Gansner2000}
Gansner,~E.~R.;\ \ North,~S.~C.  An Open Graph Visualization System and Its
  Applications to Software Engineering,  \textit{Softw: Pract. Exper.}
  \textbf{2000,} \textsl{30,} 1203--1233.

\end{thebibliography}

\providecommand{\refin}[1]{\\ \textbf{Referenced in:} #1}

\end{document}